\documentclass[12pt,a4paper]{article}
\usepackage{amsmath,amssymb}
\usepackage{graphicx}
\usepackage{cite}

\allowdisplaybreaks[3]

\begin{document}

\begin{titlepage}

\begin{flushright}
KEK-TH-2142, NCTS-TH-1906
\end{flushright}

\vspace{3em}

\begin{center}
{\Large\textbf{de Sitter duality and\\logarithmic decay of dark energy}}
\end{center}

\begin{center}
Hiroyuki \textsc{Kitamoto}$^{1)}$
\footnote{kitamoto@cts.nthu.edu.tw}
Yoshihisa \textsc{Kitazawa}$^{2),3)}$
\footnote{kitazawa@post.kek.jp} 
Takahiko \textsc{Matsubara}$^{2),3)}$
\footnote{tmats@post.kek.jp} 
\end{center}

\begin{center}
$^{1)}$
\textit{Physics Division, National Center for Theoretical Sciences}\\
\textit{National Tsing-Hua University, Hsinchu 30013, Taiwan}\\
$^{2)}$
\textit{KEK Theory Center, Tsukuba, Ibaraki 305-0801, Japan}\\
$^{3)}$
\textit{Department of Particle and Nuclear Physics}\\
\textit{The Graduate University for Advanced Studies (Sokendai)}\\
\textit{Tsukuba, Ibaraki 305-0801, Japan}
\end{center}

\begin{abstract}
We investigate infrared dynamics of four-dimensional Einstein gravity in de Sitter space. 
We set up a general framework to investigate dynamical scaling relations 
in quantum/classical gravitational theories. 
The conformal mode dependence of Einstein gravity is renormalized to the extent 
that general covariance is not manifest. 
We point out that the introduction of an inflaton is necessary as a counterterm. 
We observe and postulate a duality between quantum effects in Einstein gravity 
and classical evolutions in an inflation (or quintessence) model. 
The effective action of Einstein gravity can be constructed 
as an inflation model with manifest general covariance. 
We show that $g=G_N H^2/\pi$: 
the only dimensionless coupling of the Hubble parameter $H^2$ 
and the Newton's coupling $G_N$ in Einstein gravity 
is screened by the infrared fluctuations of the conformal mode. 
We evaluate the one-loop $\beta$ function of $g$ with respect to the cosmic time $\log Ht$ 
as $\beta(g)=-(1/2)g^2$, i.e., $g$ is asymptotically free toward the future. 
The exact $\beta$ function with the backreaction of $g$ reveals the existence of the ultraviolet fixed point. 
It indicates that the de Sitter expansion stared at the Planck scale with a minimal entropy $S=2$. 
We have identified the de Sitter entropy $1/g$ with the von Neumann entropy of the conformal zero mode. 
The former evolves according to the screening of $g$ and the Gibbons-Hawking formula. 
The latter is found to increase by diffusion in the stochastic process at the horizon in a consistent way. 
Our Universe is located very close to the fixed point $g=0$ with a large entropy. 
We discuss possible physical implications of our results such as logarithmic decay of dark energy. 

\end{abstract}

\vspace{\fill}

Jan. 2020

\end{titlepage}

\section{Introduction}
\setcounter{equation}{0}

In de Sitter-type spaces,  nontrivial scaling phenomena have been observed. 
de Sitter space is scale invariant while more nontrivial scaling laws hold 
in the temperature fluctuations of the cosmic microwave background (CMB). 
It is very desirable to determine the equation of state $w$ for dark energy. 
de Sitter space is the solution of the Einstein equation 
with a positive cosmological constant or the Hubble parameter $H$. 
It may exhibit a nontrivial dynamical scaling behavior at quantum level. 
The infrared (IR) behavior of Einstein gravity in de Sitter space is likely to be highly nontrivial 
as it has an event horizon. 
The smallness of $H^2$ in comparison with the Newton's coupling $G_N$, 
i.e., the smallness of the dimensionless coupling $g=G_NH^2/\pi$, is a quintessential problem. 
The other side of the coin is to explain the hugeness of the de Sitter entropy $S=1/g$. 
It is very desirable to find out what carries such huge entropy. 

Nontrivial scaling laws are easy to implement in slow-roll inflation theories with various inflaton potentials. 
The problem here is the embarrassment of riches. 
There are too many inflation models as we lack a principle to constrain them. 
We have formulated a duality between quantum and classical gravitational theories in two dimensions 
as Liouville gravity/inflation theory duality \cite{KK2d}. 
In this paper, we argue that such a concept of duality works equally well in four dimensions. 
It may be regarded as constructing an effective action of Einstein gravity by an inflation theory. 
We are concerned with quantum IR effects due to the presence of the horizon. 

The history of seeking a mechanism to screen the cosmological constant is long 
\cite{Polyakov1,Polyakov2,Jackiw}. 
The essential feature of our mechanism is the diffusion of the conformal zero mode 
and the creation of entropy. 
The negative metric of the conformal mode is crucial for screening (negative anomalous dimension) 
of the cosmological constant operator \cite{KK,BH}. 
In our mechanism, the IR logarithmic effects play an essential role \cite{TW,Weinberg}. 
We evaluate the one-loop dynamical $\beta$ function of $g$ with respect to the cosmological time $\log Ht$ 
to confirm the screening effects: $\beta(g)=-(1/2)g^2$. 
The negative sign implies that $g$ is asymptotically free toward the future \cite{GW,PT}. 
Our interpretation of the de Sitter entropy as a von Neumann entropy is consistent 
with the $\beta(g)$ function in four-dimensional de Sitter space. 
We have built on the stochastic picture of IR fluctuations \cite{STJY,Woodard}. 
We show that the de Sitter entropy is created at the horizon by a diffusion 
and it reduces the cosmological constant in a consistent way with $\beta(g)$. 
The dual picture to account for the increase of the de Sitter entropy in inflation theory is 
the incoming inflaton energy flux \cite{FK}. 

We also derive the exact $\beta$ function within the Gaussian approximation 
by taking into account the backreaction of $g$. 
The exact $\beta$ function is negative in the whole region of time flow. 
Furthermore, it possesses the ultraviolet (UV) fixed point in the past $g=1/2$. 
This fact indicates that our Universe started the de Sitter expansion with a minimal entropy $S=2$ 
while it has $S=10^{120}$ now. 

We believe that our results are universal, i.e., independent of the microscopic theory of quantum gravity. 
Of course, the construction of de Sitter space in string theory is a challenging task \cite{KKLT,Vafa}. 
Nevertheless, the investigation of quantum IR effects in de Sitter-type spaces is necessary 
to unlock the secrets of the Universe. 

We focus on quantum IR effects which are characteristic to de Sitter space. 
Due to the scale invariant spectrum, 
the two-point function of the massless minimally coupled modes exhibits 
the logarithmic growth with time: $\log a_c,\ a_c=e^{Ht}$. 
We sum up these IR logarithmic effects by using the technique of the renormalization group. 
Since $g$ is very small even at the inflation epoch, 
the Gaussian approximation should be very good. 
In this sense, we have done the most important work. 
We sum up all leading IR effects $\log^n a_c=(Ht)^n$ to the one-loop order. 
It is essential to understand global (long term) evolution of the Universe. 
Our IR cutoff is the size of the Universe which acts as the low momentum cutoff. 
We are interested in the large $a_c$ limit which corresponds to the removal of IR cutoff. 
The determination of the $\beta$ function and the existence of a future fixed point at $g=0$ implies 
that the existence of the $a_c \rightarrow \infty$ limit. 
Fortunately, it turns out to be flat spacetime rather than de Sitter space. 

This paper consists of the following sections and appendices. 
This first section is devoted to the introduction. 
In Sec. 2, we investigate dynamical scaling laws in 4D de Sitter-type spaces. 
We argue that a duality is the key to reconcile quantum effects and general covariance. 
In Sec. 3, we investigate quantum IR effects in 4D de Sitter spaces. 
We argue that an inflaton is necessary as a covariant counterterm. 
We show that the Hubble parameter is screened by IR logarithmic effects of the conformal mode. 
In Sec. 4, we investigate the de Sitter entropy.
We confirm that it increases in a consistent manner with the Gibbons-Hawking formula $S=\pi/(G_N H^2)$. 
We sum up leading IR logarithms by a Fokker-Planck equation. 
We derive the $\beta$ function for $g=1/S$ and find that $g$ decays logarithmically toward the future. 
This is the most important result of the paper and may have deep implications. 
We discuss some of them such as logarithmic decay of dark energy in Sec. 5. 
We compare the predictions of our theory and the standard $\Lambda$CDM model 
with the recent observations of dark energy. 
Our theory has characteristic features and it fares well with the $\Lambda$CDM model. 
We are convinced that the difference is observable in the near future. 
We conclude with discussions in Sec. 6. 
In Appendix A, we recall our propagators in a Becchi-Rouet-Stora-Tyutin gauge fixing for self-containedness. 
In Appendix B, we explain a duality between quantum effects in Einstein gravity and inflation theory in detail. 

\section{Duality and scaling in 4D de Sitter space}
\setcounter{equation}{0}

In this section, we study dynamical scaling laws in 4D de Sitter-type gravity. 
The quantum gravity is such an example while an inflation theory is another \cite{Infst,InfAG,InfLD,InfPS}. 
We seek a generic framework to encompass them. 
Our working assumption is that there is a duality between a quantum gravity and an inflation theory. 
For example, the quantum effects of Einstein gravity can be reproduced 
as a classical solution by an inflation theory. 
We may call it quantum gravity/inflation theory duality. 
 
Our duality is based on the fact that Einstein gravity is likely to be renormalized 
beyond recognition by quantum IR effects. 
We show that manifest general covariance is lost at the one-loop level. 
It is because the tree action does not admit nontrivial scaling laws. 
We thus need a practical method to ensure general covariance on the effective action. 
In two-dimensional gravity, the conformal invariance provides such a tool. 
We claim that manifest general covariance can be kept in a dual inflation theory. 
On the other hand, the duality puts discipline on the inflation theory. 
Einstein gravity possesses the shift symmetry in the weak coupling limit 
since de Sitter space has a flat potential, i.e., the cosmological constant. 
Inflation theory may be regarded as a low energy effective theory of Einstein gravity. 
Such a duality may hold only at the beginning of the inflation. 
Afterward, the inflation theory may evolve by its own logic such as QED or QCD. 
As Einstein gravity is a very good description of the current Universe, 
this duality may be applicable to dark energy and quintessence theory \cite{RP,CDS}. 
In this context, we may call it quantum gravity/quintessence duality. 

As for the principle driving force of the quantum IR corrections in Einstein gravity, 
we focus on the scale invariant fluctuations of the metric, especially the conformal mode. 
It causes logarithmic growth of quantum gravitational corrections. 
In a stochastic picture, zero modes perform a Brownian motion by the collisions with newcomers 
in field space (real line) since the two-point function at the coincident point grows linearly with cosmic time. 

In dealing with the quantum fluctuations whose background is de Sitter space, 
we adopt the following parametrization: 
\begin{align}
g_{\mu\nu}=\Omega^2(x)\tilde{g}_{\mu\nu},\hspace{1em}
\Omega(x)=a(\tau)\phi(x),\hspace{1em}
\phi(x)=e^{\omega(x)}, 
\label{para1}\end{align}
\begin{align}
\det \tilde{g}_{\mu\nu}=-1,\hspace{1em}
\tilde{g}_{\mu\nu}=\eta_{\mu\rho}(e^{h(x)})^\rho_{\ \nu}=(e^{h(x)})_\mu^{\ \rho}\eta_{\rho\nu}. 
\label{para2}\end{align}
The inverse metric  matrix is 
\begin{align}
\tilde{g}^{\mu\nu}=(e^{-h(x)})^\mu_{\ \rho}\eta^{\rho\nu}=\eta^{\mu\rho}(e^{-h(x)})_\rho^{\ \nu}. 
\label{parainv}\end{align}
To satisfy (\ref{para2}), $h_{\mu\nu}$ is traceless 
\begin{align}
\eta^{\mu\nu}h_{\mu\nu}=0. 
\label{para3}\end{align}

By using this parametrization, the components of the Einstein-Hilbert action are written as follows. 
We keep a parameter $D$ to specify the dimension for generality: 
\begin{align}
\sqrt{-g}=\Omega^D, 
\label{com1}\end{align}
\begin{align}
R=\Omega^{-2}\tilde{R}-2(D-1)\Omega^{-3}\tilde{g}^{\mu\nu}\nabla_\mu\partial_\nu\Omega
-(D-1)(D-4)\Omega^{-4}\tilde{g}^{\mu\nu}\partial_\mu\Omega\partial_\nu\Omega, 
\label{com2}\end{align}
where $\tilde{R}$ is the Ricci scalar constructed from $\tilde{g}_{\mu\nu}$, 
\begin{align}
\tilde{R}=-\partial_\mu\partial_\nu\tilde{g}^{\mu\nu}
-\frac{1}{4}\tilde{g}^{\mu\nu}\tilde{g}^{\rho\sigma}\tilde{g}^{\alpha\beta}\partial_\mu\tilde{g}_{\rho\alpha}\partial_\nu\tilde{g}_{\sigma\beta}
+\frac{1}{2}\tilde{g}^{\mu\nu}\tilde{g}^{\rho\sigma}\tilde{g}^{\alpha\beta}\partial_\mu\tilde{g}_{\sigma\alpha}\partial_\rho\tilde{g}_{\nu\beta}. 
\label{com3}\end{align}

From (\ref{com1}) and (\ref{com2}), the Lagrangian of Einstein gravity is 
\begin{align}
&\frac{1}{\kappa^2}\int d^D x\sqrt{-g}\big[R-(D-1)(D-2)H^2\big] \notag\\
=&\frac{1}{\kappa^2}\int d^4 x
\big[\Omega^2\tilde{R}-6\Omega\partial_\mu (\tilde{g}^{\mu\nu} \partial_\nu\Omega)
-6H^2\Omega^4\big] \notag\\
=&\frac{1}{\kappa^2}\int d^4 x
\big[\Omega^2\tilde{R}+6\tilde{g}^{\mu\nu}\partial_\mu\Omega \partial_\nu\Omega
-6H^2\Omega^4\big], 
\label{gravity}\end{align}
where $\kappa$ is defined by the Newton's coupling $G_N$ as $\kappa^2=16\pi G_N$. 
In the last equality, we dropped a total derivative term.
However, this operation changes the value of the action from $6H^2$ to $-12H^2$ when $D=4$. 
The former has the geometric expression with the correct semiclassical de Sitter entropy.

In the conformally flat coordinate, i.e., Poincar\'e patch, the equations of motion are
\begin{align}
ds^2=a^2(-d\tau^2+dx_i^2), 
\end{align}
\begin{align}
a:\ \partial_0^2 a=2H^2a^3, 
\label{Eq1}\end{align}
\begin{align}
h^{00}:\ \partial_0^2 a^2=6\partial_0 a\partial_0 a. 
\label{EQMT}\end{align}

Four-dimensional de Sitter space is the solution of both equations: 
\begin{align}
a_c=\frac{1}{-H\tau}=e^{\omega_c}, 
\end{align}
\begin{align}
ds^2=\big(\frac{1}{-H\tau}\big)^2 (-d\tau^2 +dx_i^2)=-dt^2+e^{2Ht}dx_i^2.  
\label{dSmetric}\end{align}
It is a maximally symmetric space $R=12H^2$ with  the action, 
\begin{align}
S=\frac{1}{\kappa^2}\int d^4 x \sqrt{g}(R-6H^2)=\frac{1}{\kappa^2}\int d^4 x \sqrt{g}\ 6H^2. 
\end{align}

From the action, the potential term is obtained for slowly varying $\phi$, 
\begin{align}
\frac{i}{16\pi G_N}\int d^4x\sqrt{-g}\ 6H^2(\phi^4-2\phi^2)
\ \rightarrow\ \frac{\pi}{G_N H^2} (\phi^4-2\phi^2), 
\end{align}
where we have compactified four-dimensional de Sitter space into $S^4$ of radius $1/H$.
The first term and the second term corresponds to $\sqrt{g}$ and $\sqrt{g}R$ respectively. 
The semiclassical formula for the geometric entropy for $dS^4$ is obtained 
at the minimum of the potential with $\phi=1$, 
\begin{align}
\frac{\pi}{G_NH^2}. 
\end{align}

Suppose the cosmological constant evolves with time while the Newton's coupling is held constant: 
\begin{align}
H^2(\tau)\propto H^2\big(\frac{1}{-H\tau}\big)^{-2\gamma}. 
\label{Hubscl}
\end{align}
According to (\ref{Eq1}), the scale of the Universe evolves as
\begin{align}
a=\big(\frac{1}{-H\tau}\big)^{1+\gamma}= \frac{1}{-H(\tau) \tau},\hspace{1em}
a H(\tau)=a_c H. 
\label{OmHR}\end{align}
We introduce the cosmic time $t$, 
\begin{align}
Ht= \frac{1}{\gamma}\big(\frac{1}{-H\tau}\big)^\gamma. 
\end{align}
The scale factor is
\begin{align}
a=(\gamma Ht)^\frac{1+\gamma}{\gamma}. 
\end{align}
The Hubble parameter decays inverse-proportionally with the cosmic time: 
\begin{align}
H(t)=\frac{\dot{a}}{a}=\frac{1}{\gamma t},\hspace{1em}
\log a=\frac{1+\gamma}{\gamma}\log (1+\gamma Ht)\sim(1+\gamma)Ht. 
\label{Hubble1}\end{align}
The $\dot{O}$ denotes the derivative with respect to the cosmic time $t$ 
such as $\dot{a}=\partial a/\partial t$. 
Note that this solution does not satisfy the other equation of motion with respect to $h^{00}$ (\ref{EQMT}) 
unless $\gamma=0$ just like 2D gravity.  

This is a serious problem which needs to be addressed in order to investigate 
possible time dependence of the cosmological constant in Einstein gravity. 
Of course, such a nontrivial solution extremizes the  effective action not the tree action. 
However, the Einstein-Hilbert action is likely to be renormalized by quantum IR effects beyond recognition. 
It may even contain new degrees of freedom.
In two dimensions, an analogous problem led us to introduce an inflaton  
as a dual description of  Liouville gravity \cite{KK2d}. 
A dual model is constructed in such a way that the classical evolution of an inflaton accounts for 
the quantum effects of Liouville gravity. 
We adopt the analogous strategy here and introduce an inflaton 
to satisfy the equation of motion with respect to $h^{00}$. 
Furthermore, its role is to provide a dual description of four-dimensional Einstein gravity. 
Namely, we adopt the inflaton potential in such a way that the classical evolution of the inflaton reproduces 
the quantum IR effects of Einstein gravity. 

As a concrete ansatz, we postulate the following Lagrangian of a single-field inflation model 
as a dual to Einstein gravity in four-dimensional de Sitter space: 
\begin{align}
\frac{1}{\kappa^2}\int d^4 x\sqrt{-g}\big[{R}
-6H^2(\gamma)\exp(-2\Gamma(\gamma )f) 
-2\Gamma(\gamma)g^{\mu\nu}\partial_\mu f \partial_\nu f\big]. 
\end{align}
It is clear from this Lagrangian that the inflaton $f$ rolls down an exponential potential. 
The Hubble parameter decreases as the Universe evolves and it eventually vanishes. 
So our proposal is a de Sitter duality between quantum and classical gravitational theories. 
This action looks as follows if we make the conformal mode $a$ dependence explicit: 
\begin{align}
\frac{1}{\kappa^2}\int d^4 x\big[a^2\tilde{R} 
+6\tilde{g}^{\mu\nu}\partial_\mu a\partial_\nu a 
-6H^2(\gamma)a^4\exp(-2\Gamma(\gamma )f) 
-2\Gamma(\gamma)a^2\tilde{g}^{\mu\nu}\partial_\mu f \partial_\nu f\big], 
\label{Inflaton}\end{align}
where $H^2(\gamma)=H^2(1+\cdots)$ and $\Gamma (\gamma) = \gamma(1+\cdots)$ 
are expanded in $\gamma$. 

The equations of motion are 
\begin{align}
a:\ -6 \partial_0^2a+12H^2(\gamma)a^3\exp(-2\Gamma(\gamma) f) 
= 2\Gamma(\gamma)a\partial_0 f \partial_0 f, 
\end{align}
\begin{align}
f:\ -4\partial_0(a^2 \partial_0f)+12H^2(\gamma)a^4\exp(-2\Gamma f)=0, 
\end{align}
where we put $\tilde{g}^{\mu\nu}=\eta^{\mu\nu}$ and then $\tilde{R}=0$. 

The equation of motion with respect to $h^{00}$ is
\begin{align}
6\partial_0a\partial_0a-\partial_0^2a^2
=2\Gamma(\gamma )a^2\partial_0 f \partial_0 f. 
\label{Eqh0}\end{align}
The solution is postulated to be
\begin{align}
a=e^f=a_c^{1+\gamma}. 
\label{EQCD}\end{align}

The Eq. (\ref{Eqh0}) is not independent as it follows from the other two equations. 
It implies 
\begin{align}
2\gamma(1+\gamma)a_c^{4+2\gamma}=2(1+\gamma)^2\Gamma a_c^{4+2\gamma}. 
\end{align}
The contribution from the inflaton fills the right-hand side of the equation. 
In fact, the two coefficients, i.e., the Hubble parameter $H^2 (\gamma)$ 
and the anomalous dimension $\Gamma(\gamma) $ can be adjusted in a simple way as follows
to establish the validity of the solution (\ref{EQCD}) to  all orders in $\gamma$: 
\begin{align}
H^2(\gamma)=H^2 (1+\frac{2}{3}\gamma)(1+\gamma),\hspace{1em}
\Gamma(\gamma) = \frac{\gamma}{1+\gamma}. 
\end{align} 

We may sweep the inflaton under the rug by using its identity with the conformal mode (\ref{EQCD}) 
in the action (\ref{Inflaton}), 
\begin{align}
\frac{1}{\kappa^2}\int d^4 x\big[a^2\tilde{R} 
+(6-2\Gamma)\tilde{g}^{\mu\nu}\partial_\mu a \partial_\nu a 
-6H^2(\gamma)a^{4(1-\frac{\Gamma}{2})} \big]. 
\label{CMact}\end{align}
The solution $a=a_c^{1+\gamma}$ also extremizes this restricted action 
as it does so in an extended field space with an inflaton. 
In this Lagrangian, the nontrivial scaling dimension of the Hubble parameter 
$H^2(t)\sim \exp(-2\Gamma f)=\exp(-2\gamma\omega_c)$ is manifest. 
The equation of motion with respect to $h^{00}$ is satisfied by the construction. 
It requires us to introduce a new counterterm. 
It is a finite renormalization of the kinetic term of the conformal mode. 
Although it is no longer manifest here, general covariance is kept intact in its dual inflation theory. 

Our earlier investigation indicated 
that the one-loop IR logarithmic corrections in Einstein gravity are of the form \cite{BH}: 
\begin{align}
\delta \frac{H^2(t)}{\kappa^2(t)}=\frac{H^2}{\kappa^2}(-4\gamma \log a_c),\hspace{1em}
\delta\frac{1}{\kappa^2(t)}=\frac{1}{\kappa^2}(-2\gamma \log a_c),\hspace{1em}
\gamma=\frac{3}{8}\frac{\kappa^2H^2}{4\pi^2}. 
\label{1loop2}\end{align}
As is explained in the next section, 
a further rescaling is necessary to fix the gravitational coupling $\kappa^2$. 

To the leading order, the quantum correction to the Hubble parameter is 
\begin{align}
H^2(t)=H^2(1-2\gamma \log a_c). 
\label{QCdual}\end{align}
It depends on the scale of the Universe $\log a_c=Ht=\omega_c$ due to IR logarithmic effects. 
This behavior (\ref{QCdual}) is consistent with our power law working hypothesis $H^2(t)\sim a_c^{-2\gamma}$ 
to the one-loop order. 
This screening effect takes place due to the accumulation of scale invariant fluctuations 
of the conformal degrees of metric. 
The screening occurs due to the negative sign of the conformal mode propagator.
These features are in common with two-dimensional Liouville gravity in the semiclassical regime. 

Our prescription to construct the dual model is to describe the quantum effects of Einstein gravity 
by the classical evolution of an inflaton: 
\begin{align}
\exp(-2\gamma\omega_c)= \exp(-2\Gamma f). 
\end{align}
We have introduced an exponential potential $\exp(-2\Gamma f)$ of the inflaton for this purpose. 
In order to cancel the IR logarithmic corrections to the Newton's coupling (\ref{1loop2}),
we rescale $a\to a e^{\gamma\omega_c}$, 
\begin{align}
\frac{1}{\kappa^2}\int d^4 x e^{2\gamma\omega_c}\big[a^2\tilde{R}
+(6-2\Gamma)\tilde{g}^{\mu\nu}\partial_\mu a \partial_\nu a -6H^2(\gamma)a^4 \big]. 
\end{align}
We note that the action acquires an overall factor $e^{2\gamma\omega_c}$ 
after this procedure which can be associated with the Newton's coupling. 
It serves as the counterterm to the Newton's coupling 
such that the physical Newton's coupling $\kappa^2(t)/a_c^{2\gamma}$ is constant. 
We have thus constructed a framework to accommodate 
a nontrivial scaling dimension of the cosmological constant operator $1-\Gamma/2=\alpha$ in Einstein gravity 
by invoking its dual inflation theory. 

In conclusion, we have constructed an inflation theory with the following scaling law: 
\begin{align}
H^2(t)\sim a_c^{-2\gamma},\hspace{1em}
a^2= a_c^{2(1+\gamma)},\hspace{1em}
\kappa^2 = \kappa^2(t)/a_c^{2\gamma} = \text{const}. 
\label{scaling}\end{align}
By this approach, we are ready to explore the dynamical scaling relations (\ref{scaling}) 
in Einstein gravity and the dual inflation theory.

\section{Quantum IR effects in 4D de Sitter space}
\setcounter{equation}{0}

As is well known, the gravitational theory has a conformal invariance for its consistency. 
In fact, the Einstein-Hilbert action can be expressed in a manifestly conformally invariant manner, 
\begin{align}
\frac{1}{\kappa^2}\int d^4x \sqrt{-g}\big[\phi^2R
+6g^{\mu\nu}\partial_\mu \phi \partial_\nu \phi-6H^2\phi^4\big]. 
\label{CMIV}\end{align}
The metric $g_{\mu\nu}$ is assumed to be conformally flat as in (\ref{dSmetric}) 
representing de Sitter space in the Poincar\'e patch. 
The conformal invariance allows us to pick a flat coordinate 
in which $\tilde{R}$ only depends on $\tilde{g}_{\mu\nu}$, 
\begin{align}
\frac{1}{\kappa^2}\int d^4x\big[\Omega^2\tilde{R}
+6\tilde{g}^{\mu\nu}\partial_\mu \Omega \partial_\nu \Omega-6H^2\Omega^4\big]. 
\end{align}

The scalar curvature transforms as follows in the conformal transformation: 
\begin{align}
R=a^{-2}\tilde{R}-6a^{-3}\partial_\mu(\tilde{g}^{\mu\nu}\partial_\nu a)=12H^2, 
\label{Ctsc}\end{align}
where the last equality holds for the de Sitter solution (\ref{dSmetric}) 
with $a=a_c,\ \tilde{g}^{\mu\nu}=\eta^{\mu\nu}$. 
The $\phi$ field corresponds to the conformal mode of the metric. 
The equation of motion for $\phi$ is readable from (\ref{CMIV}), 
\begin{align}
\frac{1}{\sqrt{-g}}\partial_\mu (\sqrt{-g}g^{\mu\nu}\partial_\nu\phi) + V'(\phi)=0. 
\label{EQMCM}\end{align}

Since the signature of the kinetic term of the conformal mode is negative, 
the potential is effectively turned upside down. 
The extremum of the potential for the conformal mode is a metastable hilltop point. 
Recall that the background $a_c$ itself is the classical solution. 
So the homogeneous solution for $\phi$ must be trivial $\phi=1$. 
As we show later that there is a flat direction on-shell 
in the extended $(\phi, h^{00})$ space along $X$ field direction. 
See Appendix A for the definition of $X$ field. 
However, such a direction is lifted in the off-shell effective action. 

Needless to say, we extremize the off-shell effective action to find a quantum solution. 
In contrast, no potential is generated in the nonlinear sigma models due to the reparametrization invariance. 
The IR logarithmic correction to the cosmological constant is highly suppressed 
in nonlinear sigma models due to the absence of the potential \cite{NLSM,KKNS}. 
On the other hand, a nontrivial potential is generated in the off-shell effective action in Einstein gravity. 
In this sense, they are totally different. 
The flatness of the potential is lifted by IR logarithmic effects at the one-loop level 
in four-dimensional Einstein gravity in de Sitter space. 

Here we explain in some detail how to evaluate the effective action with IR effects 
in a background gauge\cite{TV}. 
The relevant propagators are listed in Appendix A for self-containedness.
The essential point is that there are two types of fields. 
The massless minimally coupled modes 
and conformally coupled modes with the effective mass $m^2=2H^2$. 
Since we are interested in IR logarithmic corrections, 
we ignore the massive modes of $m^2= \mathcal{O}(H^2)$ and work in the subspace. 
Let us consider the homogeneous and isotropic background: 
\begin{align}
\hat{g}_{\mu\nu}=a^2(\tau)\eta_{\mu\nu}, 
\end{align}
where the time dependence of the scale factor is not specified 
except being close to de Sitter space with small but arbitrary perturbations. 
The Ricci tensor as shown below becomes proportional to the metric tensor on-shell which is conformally flat
\begin{align}
a^2\hat{R}_{00}=-3a\partial_0^2a+3\partial_0 a\partial_0 a,\hspace{1em}
a^2\hat{R}_{ij}=(a\partial_0^2a+\partial_0 a\partial_0 a)\delta_{ij},\hspace{1em}
a^4\hat{R}=6a\partial_0^2 a. 
\end{align}

On the general background, the quadratic action for each field is given by 
\begin{align}
&\frac{1}{\kappa^2}\int d^4x\sqrt{-g}[R-6H^2]\big|_2\notag\\
=&\frac{1}{\kappa^2}\int d^4x\sqrt{-\hat{g}}\big\{
-\frac{1}{4}\hat{g}^{\mu\nu}\nabla_\mu h^\rho_{\ \sigma}\nabla_\nu h^\sigma_{\ \rho} 
+\frac{1}{2}\hat{g}^{\mu\nu}\nabla_\rho h^\rho_{\ \mu}\nabla_\sigma h^\sigma_{\ \nu} 
+\frac{1}{2}\hat{R}^{\mu\nu}_{\ \ \rho\sigma}h^\rho_{\ \mu}h^\sigma_{\ \nu} \notag\\
&\hspace{7em}-2\hat{g}^{\mu\nu}\nabla_\rho h^\rho_{\ \mu}\partial_\nu \omega 
-2\hat{R}^\mu_{\ \nu}h^\nu_{\ \mu}\omega \notag\\
&\hspace{7em}+6\hat{g}^{\mu\nu}\partial_\mu \omega \partial_\nu \omega
+2\hat{R}\omega^2-48H^2\omega^2\big\}, 
\end{align}
\begin{align}
\int d^4x\mathcal{L}_\text{GF}&=\frac{1}{\kappa^2}\int d^4x\sqrt{-\hat{g}}
\big[-\frac{1}{2}\hat{g}^{\mu\nu}F_\mu F_\nu\big], \notag\\
F_\mu&=\nabla_\rho h^\rho_{\ \mu}-2\partial_\mu\omega
-2a^{-1}\partial_0a(h^0_{\ \mu}-2\delta^0_{\ \mu}\omega), 
\end{align}
\begin{align}
\int d^4x\mathcal{L}_\text{FP}\big|_2=
\frac{1}{\kappa^2}\int d^4x\big[-a^2\partial_\mu \bar{b}^i\partial^\mu b^i
+a^2\partial_\mu \bar{b}^0\partial^\mu b^0+(-2a\partial_0^2a+6\partial_0a\partial_0a)\bar{b}^0b^0\big]. 
\end{align}
The Lorentz indices are raised and lowered by $\eta^{\mu\nu}$ and $\eta_{\mu\nu}$ respectively 
when the scale factor $a$ is explicitly expressed. 

Our task to evaluate the one-loop IR effects in the effective action is accomplished 
just by contracting the quadratic terms. 
The Einstein-Hilbert action induces the IR logarithms as follows 
\begin{align}
&\frac{1}{\kappa^2}\int d^4 x\sqrt{-g}[R-6H^2]\big|_\text{1-loop} \notag\\
\simeq &\frac{1}{\kappa^2}\int d^4 x\big[ 
2 a\partial_0a \langle h^{0\mu}\partial_\nu h^\nu_{\ \mu}\rangle 
-8 a\partial_0a \langle h^{0\mu}\partial_\mu \omega\rangle \notag\\
&\hspace{4em}+3\partial_0a\partial_0a \langle h^{0\mu}h^0_{\ \mu}\rangle
+(4a\partial_0^2a-8\partial_0a\partial_0a) \langle h^{00}\omega\rangle \notag\\
&\hspace{4em}+(12a\partial_0^2a-48a^4)\langle\omega^2\rangle \big] \notag\\
\simeq &\frac{1}{\kappa^2}\int d^4 x\big[
24a\partial_0^2a-48H^2a^4+(8a\partial_0^2a-16\partial_0a\partial_0a)\big]\langle \omega^2\rangle. 
\end{align}
In the first line, we neglected the terms with twice-differentiated propagators 
which do not induce the IR logarithms. 
In the second line, we made use of the following identities which hold true in the subspace of massless fields: 
\begin{align}
h^{00}\simeq 2\omega,\hspace{1em}h^{0i}\simeq 0. 
\label{ids}\end{align}
We also performed partial integrations. 
In a similar way, the IR effect from the gauge fixing term is evaluated as 
\begin{align}
\int d^4x\mathcal{L}_\text{GF}\big|_\text{1-loop}
&\simeq \frac{1}{\kappa^2}\int d^4 x\big[
-2 a\partial_0a \langle h^{0\mu}\partial_\nu h^\nu_{\ \mu}\rangle 
+8 a\partial_0a \langle h^{0\mu}\partial_\mu \omega\rangle \notag\\
&\hspace{6em} -2\partial_0a\partial_0a\langle h^{0\mu}h^0_{\ \mu}\rangle
+(4a\partial_0^2a-4\partial_0a\partial_0a)\langle h^{00}\omega\rangle \notag\\
&\hspace{6em}+(4a\partial_0^2a+12\partial_0a\partial_0a)\langle\omega^2\rangle \big] \notag\\
&\simeq 0. 
\end{align}
We confirm that the gauge fixing term does not induce the IR logarithms. 
The Faddeev-Popov ghost term also does not induce the IR logarithms 
\begin{align}
\int d^4x\mathcal{L}_\text{FP}\big|_\text{1-loop}
&\simeq \frac{1}{\kappa^2}\int d^4 x 
(-2a\partial_0^2a+6\partial_0a\partial_0a)\langle \bar{b}^0b^0\rangle \notag\\
&\simeq 0. 
\end{align}
It is because $b^0$ is a massive mode, 
\begin{align}
b^0\simeq 0. 
\label{id}\end{align}

The merit of the background gauge is 
that we only need to make contractions of pairs of fields in the Einstein-Hilbert action 
to derive the one-loop effective action. 
The gauge fixing term just determines the gravitational propagators, 
and the Faddeev-Popov ghost term does not contribute to the one-loop effect. 

The one-loop effective action is obtained by simply taking the local average, 
\begin{align}
\frac{1}{\kappa^2}\int d^4x\sqrt{-\hat{g}}\big[(\hat{R} - 12H^2)\langle 4\omega^2\rangle
-2\hat{R}^\mu_{\ \nu}\langle h^\nu_{\ \mu}\omega\rangle \big]. 
\label{1lpeff3}\end{align}
Note that the effective action vanishes on-shell. 
It is because we have focused on IR logarithms and hence massless minimally coupled modes. 
Since they become exactly massless on-shell, this is what is expected.
We notice a Lorentz  symmetry breaking term (traceless symmetric tensor) 
due to the nonvanishing expectation value in our gauge: 
\begin{align} 
\langle h^{\mu\nu}\omega\rangle
\simeq-\frac{\kappa^2H^2}{4\pi^2} \log a_c
\big\{\frac{3}{8}\delta^\mu_{\ 0}\delta^\nu_{\ 0}
+\frac{1}{8}(\eta_{\mu\nu}+\delta^\mu_{\ 0}\delta^\nu_{\ 0})\big\}, 
\end{align}
\begin{align}
a^2\hat{R}_{\mu\nu}h^{\mu\nu}
\simeq a^2(\hat{R}_{00}+\hat{R}_{11})h^{00}
=(-2a\partial_0^2a +4\partial_0a\partial_0a)h^{00}, 
\end{align}
\begin{align}
a^2\hat{R}_{\mu\nu}\langle h^{\mu\nu}\omega\rangle
\simeq a^2(\hat{R}_{00}+\hat{R}_{11})\langle h^{00}\omega\rangle
\simeq 2\langle\omega^2\rangle(-2a\partial_0^2a +4\partial_0a\partial_0a). 
\label{NC2P}\end{align}
This noncovariant term also vanishes on-shell,
as it is the equation of motion with respect to $h^{00}$, 
\begin{align}
-\frac{\delta}{\delta h^{00}}\int d^4 x\sqrt{-g}R 
= 6\partial_0 a\partial_0 a-\partial_0^2a^2
= 4\partial_0 a\partial_0 a - 2 a\partial_0^2 a. 
\label{Eqh1}\end{align}
It imposes a strong constraint on the time dependence of the conformal mode $a$. 
The scale factor is determined as $a_c\propto 1/(-H\tau)$ and no other scaling is allowed. 

Nevertheless, we  explore the off-shell effective action 
as we seek a nontrivial solution with an anomalous dimension $\gamma$. 
We refrain from the shift of the Lorentz tensor $h^{00}$ to cancel this term (\ref{NC2P}) 
as it is problematic with respect to the Lorentz symmetry. 
We need to preserve it as a fundamental principle in general relativity. 
With an ansatz  $a=a_c^{1+\gamma} $ of a nontrivial dynamical scaling exponent $\gamma$, 
we find that the coefficient (\ref{Eqh1}) no longer vanishes as follows 
\begin{align}
4\partial_0 a\partial_0 a - 2 a\partial_0^2 a=2\gamma (1+\gamma)H^2a_c^{4+2\gamma}. 
\label{SGIF}\end{align}

We need to add a counterterm to subtract the right-hand side of (\ref{SGIF}) which is $\mathcal{O}(\gamma)$. 
Although the IR logarithm comes from the two-point function $\langle h^{00}\omega\rangle=-\gamma\log a_c$, 
it is necessary to cancel the $h^{00}$-tadpole first. 
Specifically, we introduce an inflaton $f$,  
\begin{align}
-2\Gamma\int d^4x \sqrt{-g}g^{\mu\nu}\partial_\mu f\partial_\nu f.  
\label{IFCT}\end{align}
We interpret this term as the $T_{00}$ component of the inflaton energy-momentum tensor 
in our construction of dual inflation theory. 
For the cancellation of the $h^{00}$-tadpole, we arrange $e^f=a_c^{1+\gamma}$, 
namely make it coincides with the conformal mode by postulating an exponential potential to $f$, 
\begin{align}
-\int d^4x  \big[\sqrt{-g} 6H^2 V(f)=6H^2a^4\exp(-2\Gamma f)=6H^2a^4a_c^{-2\gamma}\big]. 
\label{PTIF}\end{align}
It should be noted that the noncovariant term is canceled simultaneously 
as (\ref{IFCT}) includes the $e^{h^{00}}\phi^2$ operator. 

We observe that this inflaton potential contains the IR renormalization factor $a_c^{-2\gamma}$ for $H^2(t)$ identified in our previous work (\ref{QCdual}).
We thus argue that an inflaton is necessary as a covariant counterterm 
to renormalize IR logarithms of Einstein gravity. 
In this sense, the introduction of an inflaton field is analogous to an anomaly. 
As explained in (\ref{CMact}), it is equivalent to a finite modification of the kinetic term of the conformal mode 
and the cosmological constant operator 
if we eliminate the inflaton by the conformal mode using their equality 
as they satisfy the identical equations of motion. 
Although it spoils manifest general covariance, 
the general covariance holds due to the presence of the dual inflation theory. 

After establishing the renormalization procedure of the traceless tensor part, 
we move on to the analysis of the trace part. 
The effective action up to the one-loop level is 
\begin{align}
\frac{1}{\kappa^2}\int d^4x\sqrt{-\hat{g}}\big[(\hat{R} - 6H^2)
+(\hat{R} - 12H^2)\big(-\frac{3}{4}\frac{\kappa^2H^2}{4\pi^2}\log a_c\big)\big]. 
\label{1loopefac}\end{align}
Let us consider the equation of motion with respect to the conformal mode: 
\begin{align}
-\hat{g}^{\mu\nu}\frac{\delta}{\delta \hat{g}^{\mu\nu}}\big\{\sqrt{-\hat{g}}(\hat{R}-6H^2)\big\}
=\sqrt{-\hat{g}}(\hat{R}-12H^2), 
\label{TReetr}\end{align}
\begin{align}
&-\hat{g}^{\mu\nu}\frac{\delta}{\delta \hat{g}^{\mu\nu}}\big\{\sqrt{-\hat{g}}(\hat{R}-12H^2)
\big(-\frac{3}{4}\frac{\kappa^2H^2}{4\pi^2}\log a_c\big)\big\} \notag\\
=&\sqrt{-\hat{g}}(\hat{R}-24H^2) \big(-\frac{3}{4}\frac{\kappa^2H^2}{4\pi^2}\log a_c\big). 
\label{EQMSK}\end{align}
The tree action is stationary with respect to the conformal mode when $\hat{R}=12H^2$. 
However, the one-loop contribution is not so, 
indicating an instability of de Sitter solution in Einstein gravity due to IR logarithmic effects. 
In the Schwinger-Keldysh  formalism, the effective action vanishes unless we introduce different fields 
(i.e., sources) on the closed path. 
The quantum equation is free from this problem. 
Our conclusion is well defined and has a physical significance. 

What we can do is to change the scale of the metric in the classical action (\ref{TReetr}) 
to restore the balance in quantum equation, 
\begin{align}
a\to a a_c^{\gamma}.  
\label{atrns1}\end{align}
This conformal transformation changes the tree action as follows\footnote{
Here the transformation is not exact as the scalar curvature is not covariant 
under the conformal transformation. 
We will explain in Appendix B that the duality is a powerful tool to obtain an exact solution.}
\begin{align}
\frac{1}{\kappa^2}\int  d^4x a_c^{2\gamma}
\big[a^4\hat{R} -6a^4H^2a_c^{2\gamma}V(f)\big].  
\end{align}
As far as $a_c^{\gamma}$ (IR logarithm) is concerned, it comes out as the overall factor, 
\begin{align}
\frac{a_c^{2\gamma}}{\kappa^2}\int  d^4x 
\big[a^4 \hat{R} -6a^4H^2\big],  
\label{DIR6m}\end{align}
where we used (\ref{PTIF}). 

Our remaining task is to combine it with the one-loop correction in (\ref{1loopefac}). 
The result is 
\begin{align}
\frac{1}{\kappa^2}\int  d^4x 
\big[a^4\hat{R} -6a^4a_c^{-2\gamma}H^2\big]
=\frac{1}{\kappa^2}\int d^4x 
\big[a^4\hat{R} -6a^4H^2\big(1-\frac{3}{4}\frac{\kappa^2H^2}{4\pi^2}\log a_c\big)\big].  
\label{DIR7m}\end{align}
We have succeeded in constructing a new solution of quantum equation
to the leading order of IR logarithms.
It exhibits a nontrivial dynamical scaling law. 
It is certainly different from de Sitter space. 
In this Universe, a nontrivial dynamical scaling law holds with an exponent $\gamma$. 
The Newton's coupling remains constant as the conformal transformation (\ref{atrns1}) 
cancels its time evolution. 
The Hubble parameter and conformal factor of the metric scales as 
\begin{align}
\frac{a_c^{2\gamma}}{\kappa^2(t)}&=\frac{1}{\kappa^2}, \notag\\
H^2(t)&=H^2\big(1-\frac{3}{4}\frac{\kappa^2H^2}{4\pi^2}\log a_c\big)\sim a_c^{-2\gamma}, \notag\\
a^2&=a_c^2\big(1+\frac{3}{4}\frac{\kappa^2H^2}{4\pi^2}\log a_c\big)\sim a_c^{2+2\gamma}, 
\label{Betafnt}\end{align}
in agreement with the scaling arguments, (\ref{Hubscl}) and (\ref{OmHR}). 
At the one-loop level, the potential is linear rather than the exponential
as we can determine the $\mathcal{O}(\gamma)$ corrections. 
It is an inflationary universe with the slow-roll parameter $\epsilon=\gamma$ and $\eta=0$. 
A further finite renormalization of the Einstein-Hilbert action 
to make (\ref{Betafnt}) fully satisfy the quantum equation 
will be explained in Appendix B in connection with the dual inflation theory. 
We also investigate the physical property of this Universe in more detail in Sec. 5. 

After a heuristic exposition, we have shown that the following dynamical scaling relation holds 
in Einstein gravity at the one-loop level: 
\begin{align}
H^2(t)=H^2\big(1-\frac{3}{4}\frac{\kappa^2H^2}{4\pi^2}\log a_c\big)\sim a_c^{-2\gamma}. 
\label{Scaling1loop}\end{align}
It is consistent with an investigation on the dynamical scaling law (\ref{scaling}) in Einstein gravity 
with $\gamma=\frac{3}{8}\frac{\kappa^2H^2}{4\pi^2}$. 
The difficulty of revealing a nontrivial dynamical scaling relations in Einstein gravity 
stems from the fact that the Einstein-Hilbert action does not allow a modification 
of the tree level de Sitter solution with respect to the time dependence. 
Nevertheless, we believe that the nontrivial dynamical scaling relation can be realized 
in quantum Einstein gravity as the one-loop IR logarithmic corrections imply. 
The construction of such a solution is complicated as we have explained. 
It is because the effective action must be renormalized 
such that $\sqrt{-g}R$ term loses its original geometric form. 
The same is true for the cosmological constant term. 
This IR renormalization feature of Einstein gravity is analogous 
to that of two-dimensional Liouville gravity \cite{KK2d}. 
The analogous feature is pointed on UV renormalization of quantum gravity 
in ($2+\epsilon$) dimensions \cite{KKN}. 

The solution of the effective action captures the quantum effects. 
We postulate that it can be constructed as an inflation (or quintessence) model. 
The duality in anti-de Sitter (AdS) space has been very successful. 
The quantum effects in conformal field theory (CFT) has been given 
by a geometric description in AdS space. 
A possible duality in de Sitter space is an outstanding problem. 
We hope that our proposal will provide a new stimulation to this subject \cite{Maldacena,Strominger,BBM,BBM2}. 
We mention some analogies between our de Sitter duality and AdS/CFT in Sec. 5. 
Our strategy is to construct the classical dual inflation model 
which incorporates quantum IR effects of Einstein gravity in de Sitter space. 
We do not assume the exact de Sitter symmetry. 
It is shown to be logarithmically broken in the next section. 

\section{de Sitter entropy and asymptotic freedom}
\setcounter{equation}{0}

de Sitter space has a cosmological horizon. 
Gibbons and Hawking pointed out that it has a geometric entropy 
proportional to the area of the horizon \cite{GH}. 
As we have found that the Hubble parameter decreases due to quantum IR effects, 
the entropy must increase simultaneously. 
In this section, we investigate four-dimensional gravity on de Sitter space from an entropic point of view. 
In particular, we focus on our conjecture concerning the identity of the de Sitter entropy. 
In our postulate, it is the von Neumann entropy of the conformal zero mode. 
As the Universe expands at an accelerated rate, zero modes accumulate at the horizon. 
In this sense, it is a natural idea. 
Why we focus on the conformal mode? 
That is because it is the only mode which couples to the cosmological constant operator. 
In other words, it is a Lorentz scalar and does not need to be contracted with derivatives. 
In fact, the other modes are suppressed in the IR region, 
though the tensor mode $h^{00}$ includes a scale invariant spectrum. 
We believe that the Lorentz symmetry is consistent only with the conformal mode condensation. 
In other words, the other modes are excluded to contribute to the de Sitter entropy. 
We have gone so far to introduce an inflaton field 
to subtract the noncovariant quantum correction in the preceding section. 
Needless to say, the Lorentz symmetry is one of the fundamental principles on which general relativity is built. 

As is well known, the geometric entropy is equal to the effective action in quantum gravity 
as there is no energy in de Sitter space. 
A detailed investigation of the de Sitter entropy by a resummation method enables us 
to determine the counterterm. 
The bare action with the counterterm in turn enables us 
to determine the $\beta$ function of the dimensionless coupling of Einstein gravity $g=G_N H^2/\pi$. 
Since the $\beta$ function with respect to time is negative, 
Einstein gravity is asymptotically free toward the future. 
It is the most exciting discovery of this paper. 
The irony is that the scaling picture in the preceding section is superseded 
by the asymptotic freedom picture immediately after in this section. 

We consider the conformal zero mode dependence of the action: 
\begin{align}
&\frac{1}{16\pi G_N }\int \sqrt{g}(Re^{2\omega}-6H^2e^{4\omega}) \notag\\
=&\frac{1}{16\pi G_N }\int \sqrt{g} 6H^2 (2e^{2\omega}-e^{4\omega}) \notag\\
=&\frac{\pi}{G_N H^2}(2e^{2\omega}-e^{4\omega}) 
\simeq \frac{\pi}{G_N H^2}(1 -4\omega^2). 
\label{CMAC}\end{align} 
We omit the gauge fixing sector as it does not produce IR logarithms in the background gauge. 

The semiclassical de Sitter entropy $\pi/(G_N H^2)$ is obtained by rotating $dS^4$ into $S^4$. 
Since $H^2(t)\sim a_c^{-2\gamma}$ in our scheme, the de Sitter entropy increases as
\begin{align}
\frac{\pi}{G_N H^2(t)}\sim \frac{3}{2\gamma }a_c^{2\gamma}
\sim \frac{3}{2\gamma }(1+2\gamma\log a_c)
=\frac{3}{2\gamma}+3Ht. 
\label{SCENT}\end{align}
This result can be reproduced in a simple estimate as
\begin{align}
\frac{\pi}{G_N H^2}(1 -4\langle\omega^2\rangle)
\sim \frac{3}{2\gamma}(1+2\gamma\log a_c)
=\frac{3}{2\gamma}+3Ht. 
\label{Enttd}\end{align}

In the dual inflation theory picture, 
the de Sitter entropy increases due to the incoming energy flux of the inflaton. 
The increase of the entropy is estimated by the first law $T\Delta S=\Delta E$ 
where $\Delta E$ is the incoming energy flux of the inflaton. 
After translating the change of the entropy into that of the Hubble parameter by the Gibbons-Hawking formula, 
one of the Einstein equation is obtained \cite{FK}, 
\begin{align}
\dot{H}(t)=-4\pi G_N \dot{f}^2. 
\end{align}
This relation implies $\dot{S}=2\pi\epsilon/(G_N H)=3H$ 
which is consistent with (\ref{SCENT}) and (\ref{Enttd}). 
This classical picture is dual to our picture, the quantum evolution of the Hubble parameter. 

It is a fundamental question to inquire the identity of the de Sitter entropy. 
We have proposed that it is the von Neumann entropy of the conformal zero mode. 
The distribution function of the zero mode is well approximated by Gaussian, 
\begin{align}
\rho (\xi,\omega)=\frac{1}{\mathcal{N}} \exp\big({-\frac{4\xi}{g}\omega^2}\big)
=\frac{1}{\mathcal{N}}\exp\big(-\frac{6\xi}{\gamma}\omega^2\big), 
\label{StSol}\end{align}
\begin{align}
\mathcal{N}=\sqrt{\frac{\pi g}{4\xi}}, 
\end{align}
where $g=G_N H^2/\pi=2\gamma/3$ denotes the inverse de Sitter entropy. 
The Gaussian approximation must be excellent since $g$ is very small. 
We have introduced a new parameter $\xi$ to control the diffusion process of the distribution. 
The distribution is diffused as $\xi$ gets smaller. 
The von Neumann entropy of the distribution grows at the same time: 
\begin{align}
S&=-\text{tr}(\rho \log \rho) \notag\\
&= \text{tr} \big\{\rho \big(\frac{4\xi}{g}\omega^2+\log \mathcal{N}\big)\big\} \notag\\
&=\frac{1}{2}(1+\log \pi g-\log 4\xi).
\label{vNen}\end{align}
Our hypothesis is that the von Neumann entropy accounts for the time dependent part of the de Sitter entropy. 
It cannot explain the initial value. 
We fix the parameter $g$ as the initial de Sitter entropy 
and let $\xi$ evolve according to a Fokker-Planck equation. 
To the leading order in the $\log a_c =Ht$ expansion, its growing speed is expected as follows 
\begin{align} 
\dot{S}=\frac{1}{2}\big(-\frac{\dot{\xi}}{\xi}\big)= 3H, 
\label{vNentp}\end{align}
to be consistent with semiclassical result (\ref{SCENT}).  

Although we have analyzed the one-loop quantum effects in the preceding section, 
there is a resummation method of the leading IR logarithms by a Fokker-Planck equation. 
The solution of the Fokker-Planck equation shows that the leading IR logarithms are power series in $Ht$ 
not $\gamma Ht$ in the Gaussian approximation. 
We thus obtain the one-loop exact result by resummation. 

The Fokker-Planck equation of the conformal zero mode is given by 
\begin{align}
\dot{\xi}\frac{\partial}{\partial \xi}\rho 
=\frac{\gamma}{2}\cdot\frac{H}{2}\frac{\partial^2}{\partial \omega^2}\rho. 
\label{GRFP}\end{align}
The left-hand side can be identified as $\dot{\rho}$.  
The distribution function $\rho$ defines the correlation functions as follows 
\begin{align}
\langle \omega^n(t)\rangle=\int d\omega \rho(\xi,\omega)\omega^n, 
\end{align}
where $n$ is a positive integer. 

The factor $\gamma/2$ represents the residue of the conformal mode propagator in the IR region. 
This is the conversion factor from $\omega$ to $X$ field.  
The point is how to treat the negative sign of the kinetic term of the conformal mode. 
We might imagine that the sign of the right-hand side is flipped into the negative. 
However, the direction of time flow is not prefixed in quantum gravity. 
The sensible choice is to let it coincide with that of entropy. 
We see later that (\ref{GRFP}) leads to entropy generation. 

We also drop the drift term. 
As explained in the preceding section, the potential is flat in $X$ field direction at the tree level. 
At the one-loop level, we have eliminated the drift force by solving the quantum equation. 
In the dual picture, the inflaton moves according to the classical drift force. 
The conformal mode diffuses due to quantum IR effects. 
We should not double count quantum diffusion and classical drift as they are the same, 
i.e., dual effects. 

The distribution with $\xi=1$ represents de Sitter space (\ref{StSol}), 
\begin{align}
\rho(\omega)\propto\exp\big\{-V(\omega)\big\},\hspace{1em}
V(\omega)=\frac{24H^2}{\kappa^2}\omega^2\Theta, 
\label{Stsol1}\end{align}
where $\Theta=8\pi^2/(3H^4)$ is the volume of $S^4$. 
It may represent an initial state of the Universe when the inflation began. 
Our following solution (\ref{Stsol2}) is a one-parameter extension of the de Sitter solution in (\ref{Stsol1}), 
\begin{align}
\rho(\xi,\omega)\propto\exp\big\{- V(\xi,\omega)\big\},\hspace{1em}
V(\xi,\omega)=\frac{24H^2}{\kappa^2}\xi\omega^2\Theta. 
\label{Stsol2}\end{align}
In fact, there is an instability of the de Sitter solution against diffusion. 
Namely, a broader distribution with decreasing $\xi$ has a larger von Neumann entropy. 

First, we obtain an equation for $\xi$ from the Fokker-Planck equation.
In the Gaussian approximation, the Fokker-Planck equation becomes 
\begin{align}
\dot{\xi}\frac{\partial}{\partial \xi}\rho=\dot{\xi}\big(\frac{1}{2\xi}\rho-\frac{4}{g}\omega^2\rho\big), 
\end{align}
\begin{align}
\frac{\gamma H}{4}\frac{\partial^2}{\partial \omega^2}\rho=-3H\xi \rho+3H\xi^2\frac{8}{g}\omega^2\rho. 
\end{align}
We obtain the equation of our target: 
\begin{align}
\dot{\xi}=-6H\xi^2. 
\label{Fpds}\end{align}
The solution is
\begin{align}
\xi=\frac{1}{1+6Ht}. 
\label{xisol}\end{align}

The von Neumann entropy is in agreement with (\ref{vNen}) to the leading order in the $Ht$ expansion, 
\begin{align}
S=-\frac{1}{2}\log \xi. 
\end{align}
The entropy generation speed is 
\begin{align}
\dot{S}=-\text{tr}(\dot{\rho} \log \rho) = 3H\xi. 
\end{align}
We have accomplished the resummation of $(Ht)^n$ to all orders. 
The von Neumann entropy always increases under the evolution of the Fokker-Planck equation, 
\begin{align}
\dot{S}
&=-\text{tr}(\dot{\rho} \log \rho) \notag\\ 
&=-\frac{3}{16}H\frac{\kappa^2H^2}{8\pi^2}
\text{tr}\big(\frac{\partial^2}{\partial \omega^2}\rho\log \rho\big) \notag\\
&=\frac{3}{16}H\frac{\kappa^2H^2}{8\pi^2}
\text{tr}\big\{\big(\frac{\partial}{\partial \omega}\rho\big)^2\frac{1}{\rho}\big\}\notag\\
&= 3H\xi.
\label{Entic}\end{align}
We have reproduced the time dependent part of the de Sitter entropy (\ref{SCENT}) 
from the Fokker-Planck equation. 
Since $\xi=1/(1+6Ht)$ is positive, this solution is entropically more favored than the de Sitter solution. 
The Fokker-Planck equation has been reduced to a diffusion equation in the Gaussian approximation.
Our results correspond to the fact that the standard deviation of the distribution increases with time 
as $(1+6Ht)$. 

It is also possible to estimate the entropy directly from the partition function of Einstein gravity. 
The partition function of the conformal zero mode sector may be rotated into $S^4$
by assuming the system is quasiequilibrium, 
\begin{align}
Z(t)&=\int d\omega e^{\frac{1}{g}(1 -4\xi(t)\omega^2)} \notag\\
&=e^\frac{1}{g}\int d \omega e^{-\frac{4}{g }\xi(t)\omega^2} \notag\\
&=e^\frac{1}{g}\sqrt{\frac{\pi g}{4\xi(t)}}. 
\label{PFEG}\end{align}
We obtain the de Sitter entropy $S(t)=\log Z(t)$ as there is no energy in de Sitter space, 
\begin{align}
S(t)=\frac{1}{g}+\frac{1}{2}\big(\log \pi g-\log 4 \xi(t)\big). 
\label{onelpa}\end{align}
It is manifest that the conformal zero mode integration gives rise to $-(1/2)\log\xi$
by exponentiating the one-loop determinant. 
The correlation functions of this theory are defined as
\begin{align}
\langle\omega^n(t)\rangle=\frac{1}{Z(t)}\int d\omega e^{\frac{1}{g}(1 -4\xi(t)\omega^2)}\omega^n. 
\end{align}
With the choice of $\xi(t)$ as (\ref{xisol}), they satisfy the identical Fokker-Planck equation.
This argument proves that the von Neumann entropy (\ref{vNentp}) reproduces 
the time dependence of the de Sitter entropy (\ref{onelpa}) in the conformal zero mode sector. 
Here we have completed a large circle to the original Euclidean gravity approach by Gibbons and Hawking. 

The Fokker-Planck equation enables us to exactly determine the one-loop IR logarithmic correction 
to the entropy, i.e., the action. 
In what follows, we use a renormalization group technique to keep track of IR logarithmic corrections. 
We define a bare action with a counterterm 
to cancel the time dependent IR correction at the one-loop level. 
We minimally remove the time dependent part as follows 
\begin{align}
S_B= \frac{1}{g(t)}+\frac{1}{2}\log  \xi(t). 
\label{onelpa'}\end{align}
Since $S_B$ is the bare action, we derive the $\beta$ functions in a standard way, 
i.e., by requiring $S_B$ to be time independent, 
\begin{align}
\beta (g)=-\frac{1}{2}g^2,\hspace{1em}\beta(g)\equiv \frac{\partial}{\partial \log (1+6Ht)}g. 
\label{onelpa1}\end{align}
Since the $\beta(g)$ function is negative, 
the coupling $g=G_N H^2/\pi$ is asymptotically free toward the future. 
It is also remarkable that this equation to determine $g$ has no small parameter. 
It indicates that we may obtain observable effects. 
On the other hand, our Universe sits very near the fixed point $\beta=g=0$ with a large entropy \cite{FN,HKK}. 
In quantum gravity, the maximal entropy principle operates 
since the entropy is directly obtained as $S=\log Z$. 
It is because quantum gravity integrates over the geometry 
and the temperature is related to the periodicity of the metric in Euclidean time direction. 

The solutions of (\ref{onelpa1}) is 
\begin{align}
\frac{1}{g(t)}&=\frac{1}{2}\log \frac{(1+6Ht)}{(1+6Ht_i)}+ \frac{1}{g_i} \notag\\
&=\frac{1}{2}\log (1+6Ht), 
\label{onelpa3}\end{align}
where $g=G_N H^2/\pi=\kappa^2H^2/(16\pi^2)$ is the dimensionless combination 
of the Hubble parameter $H^2$ and the Newton's coupling $G_N$. 
The dimensionless coupling $g(t)$ increases toward the past. 
Its initial value is given by the time $t_i$ when the de Sitter expansion started, 
\begin{align}
\frac{1}{g_i} = \frac{1}{2}\log (1+6Ht_i).  
\end{align} 
The ratio of the couplings has the simple expression: 
\begin{align}
\frac{g(t_i)}{g(t)}=\frac{\log (1+6Ht)}{\log (1+6Ht_i)}. 
\label{olp33}\end{align}
We can introduce an analog of the QCD $\Lambda$ parameter $t_{\Lambda}$ as follows 
\begin{align}
g(t)=\frac{2}{\log (1+6Ht)-\log (1+6Ht_\Lambda)}. 
\label{onelpa4}\end{align}
In our formula (\ref{onelpa3}), we have adopted the convention $t_\Lambda=0$. 
Just like QCD, the coupling $g$ becomes large at $t=t_\Lambda$. 
These solutions are globally defined from the beginning of the de Sitter expansion 
until the end of the accelerating expansion. 
We discuss the property of the solutions of the renormalization group $\beta$ function
in comparison to the de Sitter expansion in the next section. 

We have evaluated the geometric entropy at the one-loop level exactly by the Fokker-Planck equation. 
In quantum gravity in de Sitter space, the geometric entropy is equal to the effective action. 
Therefore, we can determine the counterterms from entropy. 
Einstein gravity in de Sitter space turns out to be asymptotically free toward the future 
as implied by the inverse relationship $S=1/g$. 
The $\beta$ function (\ref{onelpa1}) controls the time evolution of the spacetime. 
It could have many implications on fundamental issues in physics. 
First of all, four-dimensional de Sitter space is doomed 
and dark energy decays logarithmically with cosmic time. 

Here we mention the previous work 
which discusses the cosmological constant problem from an analogy 
between the conformal sector in Einstein gravity 
and the $\phi^4$ theory in the flat spacetime \cite{Polyakov1}. 
The flat spacetime setup focuses on the subhorizon dynamics which respects the de Sitter symmetry. 
Therefore, the cosmological constant does not acquire time dependence. 
In contrast, our work focuses on the superhorizon dynamics 
which is expressed as the stochastic procedure with the de Sitter symmetry breaking. 
The geometric entropy increases with time, and the cosmological constant decreases simultaneously. 

Let us check to what extent our estimate of the screening of $g$ in the preceding section 
can be trusted in comparison to the one-loop exact result in this section. 
The one-loop evaluation of the IR logarithmic effects (\ref{Betafnt}) is a local estimate of the $\beta$ function. 
It obeys a scaling law as follows 
\begin{align}
\frac{1}{g}\sim \frac{1}{g_0 }e^{3g_0 Ht}\sim \frac{1}{g_0}(1+3g_0Ht)=\frac{1}{g_0}+3Ht. 
\label{Lobeta}\end{align}
On the other hand, the exact one-loop $\beta$ function gives the following time evolution:  
\begin{align}
\frac{1}{g}\sim \frac{1}{2}\log (1+ 6Ht)+\frac{1}{g_0} \sim \frac{1}{g_0}+3Ht. 
\label{Glbeta}\end{align}
Their local behaviors are identical while they behave in different ways globally, i.e., $Ht>1$. 
The $g$ in (\ref{Lobeta}) decays exponentially and the resummed $g$ in (\ref{Glbeta}) decays logarithmically. 
In evaluating the effective action, we just exponentiated the linear deformation. 
On the other hand, the Fokker-Planck equation sums up all leading powers of $Ht$ 
to form a globally valid one-loop solution. 
It has revealed asymptotic freedom toward the future, i.e., the logarithmic violation of scaling. 

Before concluding this section, we comment on the gauge dependence of the $\beta$ function. 
It has been pointed out that Einstein gravity on de Sitter space screens dimensionless couplings 
of generic field theories \cite{KK}. 
The mass parameters are not renormalized presumably due to the energy conservation. 
The anomalous dimensions $\gamma_i$ of the operators $O_i$ due to IR fluctuations 
are found to be gauge dependent. 
In a generalized gauge with a gauge parameter $\delta$, 
$\gamma_i$ in the gauge of this paper becomes $(2-\delta^2)\gamma_i$. 
In the case of the $\beta$ function, the gauge dependence appears 
only through the definition of $\mathcal{T}\equiv 1+6(2-\delta^2)Ht$ in (\ref{onelpa1}). 
The $\beta$ function does not depend on the linear redefinition of $\mathcal{T}$ 
since it is defined by the derivative with respect to $\log \mathcal{T}$. 
Therefore, the $\beta$ function for $g$ is gauge independent. 

We find that the anomalous dimensions also become gauge independent 
if we assume that $\mathcal{T}$ sets the timescale: 
\begin{align}
\gamma_i Ht = \frac{\gamma_i}{6}(\mathcal{T}-1)\sim \frac{\gamma_i}{6}\log \mathcal{T},\hspace{1em}	
\Gamma_i \equiv \frac{\partial}{\partial \log \mathcal{T}} \big(\frac{\gamma_i}{6}\log \mathcal{T}\big)
=\frac{\gamma_i}{6}. 
\end{align}
The gauge independent anomalous dimensions of the couplings in the standard model are listed below 
\begin{align}
\Gamma_{e^2_i}&= -\frac{2\gamma}{3}e^2_i,\hspace{1em}\text{gauge couplings}, \notag\\
\Gamma_{Y_i}&=-\frac{13\gamma}{24}\lambda_{Y_i,}\hspace{1em}\text{Yukawa couplings}, \notag\\
\Gamma_{\lambda_4}&=-\frac{7\gamma}{3}\lambda_4,\hspace{1em}\text{Higgs coupling}. 
\end{align}
As is well known, the presence of the fixed point 
and the sign of the first derivative at the fixed point of the $\beta$ function 
is prescription independent. 
  
Remarkably, our proposal works not only in two dimensions but in four-dimensional de Sitter space as well. 
We have gathered convincing evidences to our conjecture:
The de Sitter entropy is indeed the von Neumann entropy of the conformal zero mode.
By analyzing dual pairs in four-dimensional accelerating Universe, 
the shielding mechanism of the cosmological constant 
and the identity of the de Sitter entropy have been well elucidated. 
The mechanism of entropy generation has been identified 
with the stochastic process at the cosmological horizon \cite{STJY}. 
Our research on four-dimensional de Sitter space reinforces such a line of thinking. 

\section{Physical implications}
\setcounter{equation}{0}

In this section, we explore physical implications of our findings 
on quantum/classical gravity duals in four-dimensional de Sitter space. 
The dimensionless parameter $G_N H^2/\pi$ decays logarithmically with the cosmic evolution. 
Einstein gravity in de Sitter space is asymptotically free toward the future. 
Our hypothesis is that Einstein gravity in de Sitter space is dual 
to an inflation (or quintessence) model. 
The merit to postulate quantum/classical gravity duality in de Sitter space is twofold. 
First, this duality enables us to gain an intuitive grasp on quantum IR effects in Einstein gravity. 
On the other hand, this duality puts constraints on the inflation (or quintessence) model.
The problem of inflation models is the lack of principle to determine the inflaton potential. 
Our duality suggests that it may be generated by quantum effects. 
Since Einstein gravity is expected to be valid close to Planck  scale, 
it is important to understand its quantum IR effects in de Sitter space. 
Our postulate is that the effective action of Einstein gravity is given by an inflation model.  

Let us recall the inflaton Lagrangian (\ref{Inflaton}): 
\begin{align}
&\frac{1}{\kappa^2}\int d^4 x \sqrt{-g}\big[R-6H^2(1-\sqrt{\gamma}\kappa f) 
-\frac{\kappa^2}{2}{g}^{\mu\nu}\partial_\mu f \partial_\nu f\big] \notag\\
=&\frac{1}{\kappa^2}\int d^4 x\big[a^2\tilde{R}
+6\tilde{g}^{\mu\nu}\partial_\mu a \partial_\nu a
-6H^2a^4(1-\sqrt{\gamma}\kappa f) 
-\frac{\kappa^2}{2}a^2\tilde{g}^{\mu\nu}\partial_\mu f \partial_\nu f\big], 
\label{Inflaton2}\end{align}
where we canonically renormalized $f$ field and redefine $H^2(\gamma)\rightarrow H^2$. 
At the one-loop level, the inflaton potential is linear, 
\begin{align}
V(f)=\frac{6H^2}{\kappa^2}(1-\sqrt{\gamma}\kappa f). 
\end{align}
The slow-roll parameters are
\begin{align}
\epsilon = \frac{1}{\kappa^2}\big(\frac{V'}{V}\big)^2=\gamma,\hspace{1em}
\eta=\frac{2}{\kappa^2}\frac{V''}{V}=0. 
\end{align}
So Einstein gravity in de Sitter space performs a slow-roll inflation 
due to quantum IR effects in an analogous way with two-dimensional Liouville gravity. 
Furthermore, the Hubble parameter eventually vanishes due to the linear potential. 
It is an attractive feature with respect to dark energy application. 
We have succeeded in constructing a quintessence model. 

The equation of motion for an inflaton in a slow-roll approximation is 
\begin{align}
3H\dot{f}=-V'=\frac{6H^2}{\kappa}\sqrt{\gamma},\hspace{1em}
H^2(t)/H^2= 1-\kappa\sqrt{\gamma}f=1-2\gamma Ht. 
\end{align}
As the inflaton rolls down the potential, the Hubble parameter decreases. 
In turn, the de Sitter entropy $S=\pi/(G_N H^2(t))=3/(2\gamma)$ increases,  
\begin{align}
\dot{S}=\frac{\pi}{G_N H^2}\kappa\sqrt{\gamma}\dot{f} 
=\frac{\pi}{G_N H^2}2H\gamma=3H. 
\label{dSentv}\end{align}
The expansion of the Universe is accelerating for small $\gamma$ 
as $-\dot{H}(t)/H^2(t)\sim \gamma < 1$. 

The equation of state is 
\begin{align}
w = \frac{p}{\rho}=-\big(1-\kappa^2\frac{\dot{f}^2}{6H^2}\big)=-(1-\frac{2}{3}\gamma). 
\label{eqst}\end{align}
It is consistent with the time dependence of the Hubble parameter, 
\begin{align}
\frac{H^2(t)}{H^2}= \exp{\int_a da'\frac{3(1+w)}{a'}}\sim a_c^{-2\gamma}\sim 1-2\gamma Ht. 
\label{EsHl}\end{align}
Note that $3(1+w) = 2\epsilon$ where $\epsilon=-\dot{H}(t)/H^2(t)$ is a slow-roll parameter. 
Since general relativity applies very well to the present Universe, 
the application of this quintessence theory to dark energy is very natural. 
Unfortunately, the equation of state $w$ in (\ref{eqst}) is very close to $-1$ in the quintessence model 
dual to Einstein gravity. 
Fortunately, what we have explained so far is the local evolution of the Universe. 
We need to take account of the global behavior of the Universe at a late time. 
We show that $H^2(t)$ decreases logarithmically right after it began 
recent accelerated expansion in (\ref{Omegaest}). 
When dark energy dominates, 
the equation of state becomes $3(1+w)\sim 1/(\log a_c\log (\log a_c))\sim 2\epsilon$. 
The slow-roll parameter $\epsilon$ decreases toward the future as 
\begin{align}
\epsilon= \frac{1}{2 Ht\log Ht}. 
\label{epslg}\end{align}
This is a very robust signature of the asymptotically free de Sitter gravity as we explain it shortly. 

We can reproduce the same physical prediction 
from the renormalized Einstein-Hilbert action (\ref{CMact}) in the dual picture. 
We recall the volume operator scale as 
\begin{align}
\int d^4 x (\sqrt{-g}\sim a^{4\alpha}), 
\end{align}
where $\alpha\sim 1-{\gamma}/2$ is the scaling dimension.
The scale factor is also obtained as the solution of (\ref{CMact}): 
\begin{align}
a=a_c^{1+\gamma}. 
\end{align}
The metric is given by
\begin{align}
ds^2=\big(\frac{1}{-H\tau}\big)^{2(1+{\gamma})}(-d\tau^2+dx_i^2)
=-dt^2+a^2(t)dx_i^2, 
\end{align}
where the scale factor is 
\begin{align}
a(t)=(\gamma Ht)^\frac{1+\gamma}{\gamma}. 
\end{align}

The Hubble parameter shows that the expansion of the Universe is accelerating 
$-\dot{H}(t)/H^2(t)\sim \gamma<1$ which is in agreement with an inflation picture, 
\begin{align}
a=(\gamma Ht)^\frac{1+\gamma}{\gamma},\hspace{1em}
H^2(t) \sim a_c^{-2\gamma}\sim H^2(1-2\gamma Ht). 
\label{Sclsol}\end{align}
These results are based on the one-loop IR logarithmic effect to shield $G_N H^2/\pi$. 
However, the picture changes dramatically by summing all leading IR logarithms by the Fokker-Planck equation. 
We then find the logarithmic breaking of scaling with the $\beta$ function 
for $g= G_N H^2/\pi$ in (\ref{onelpa1}). 
The scaling picture is replaced by the asymptotic freedom picture. 
The dimensionless Hubble parameter $g$ decays logarithmically with the cosmic evolution. 
It implies that dark energy also decays logarithmically. 
We will come back to this subject as a finale of this paper. 

We point out an illuminating example of the solutions of the $\beta$ function. 
Here we reparametrize $\log (1+6Ht)\rightarrow \log (1+Ht)$ 
using the invariance of the $\beta$ function under such a linear transformation. 
The scale factor $a(\tau)$ of de Sitter space can be regarded as such a solution, 
\begin{align}
g(t)=\frac{2}{\log (1+Ht)}\ \leftrightarrow\ 2a (\tau )=\frac{2}{-H\tau}. 
\end{align}
It is because $g(t)$ and $2a (\tau)$ satisfy the same equation: 
\begin{align}
\frac{\partial}{\partial (-H\tau)} 2a(\tau )=-\frac{1}{2}\big(\frac{2}{-H\tau}\big)^2
=-\frac{1}{2}(2a(\tau ))^2, 
\end{align}
under the following reparametrization of the variables: 
\begin{align}
\log (1+Ht),\ t>0\ \leftrightarrow\ -H\tau,\ \tau < 0. 
\label{CrDs}\end{align}

Another interesting property of the solution is the time reversal symmetry 
under $-H\tau \leftrightarrow 1/ (-H \tau)$. 
Namely, $-2H\tau$ satisfies the same equation: 
\begin{align}
\frac{\partial}{\partial \frac{1}{-H\tau}}(-2H\tau)
=H\tau^2\frac{\partial}{\partial \tau}(-2H\tau)
=-\frac{1}{2} (-2H\tau)^2. 
\end{align}
So the inverted function is also the solution of the $\beta$ function: 
\begin{align}
g(t)=2\log (1+Ht)\ \leftrightarrow\ 2a (\tau )=-2H\tau. 
\label{Ivml}\end{align}
This inversion corresponds to the time reversal symmetry. 
As we have pointed out in (\ref{olp33}), 
the simplest solution of the $\beta$ function is a ratio of the solutions like $H^2(t)/H^2(t_0)$: 
\begin{align}
\frac{H^2(t)}{H^2(t_0)}=\frac{\log (1+Ht_0)}{\log (1+Ht)}
\ \leftrightarrow\ \frac{\log (1+Ht)}{\log (1+Ht_0)}. 
\label{trsm}\end{align}
We can simply invert the ratio when we change the direction of time flow. 
 
In view of the transition behavior from the exponential dumping to the logarithmic dumping, 
we need to modify the potential, i.e., running $H^2(t)$, as follows: 
\begin{align}
H^2(1-2\gamma Ht)\rightarrow \frac{2H^2}{-H\tau}= \frac{2H^2}{\log (1+Ht)}\equiv H^2(t).
\label{glpot}\end{align}
The renormalized action is 
\begin{align}
\frac{1}{\kappa^2}\int d^4 x\big[a^2\tilde{R}
+(6-2\Gamma)\tilde{g}^{\mu\nu}\partial_\mu a \partial_\nu a
-6H^2(t)a^4 \big]. 
\label{CMact1}\end{align}
This is essentially the Einstein-Hilbert action with the running $H^2(t)$. 

Let us consider what is the dual inflaton theory to Einstein gravity 
including all leading one-loop IR logarithms. 
The question is what is the potential of inflaton $V(f)$ in such a theory, 
\begin{align}
\frac{1}{\kappa^2}\int d^4 x \sqrt{-g}\big[R
-6H^2V(f) -\frac{\kappa^2}{2}g^{\mu\nu}\partial_\mu f \partial_\nu f \big]. 
\end{align}

We examine the linear potential $V(f) = 1-\sqrt{\gamma}\kappa f$. 
The equation of motion in the slow-roll approximation is
\begin{align}
3H(t)\dot{f}=\frac{6H^2}{\kappa}\sqrt{\gamma}. 
\label{Nleq}\end{align}
First, let us assume that $f$ is small.
The Hubble parameter $H^2(t)/H^2$ behaves as 
\begin{align}
V(f)=1-\sqrt{\gamma}\kappa f=1-2\gamma Ht=1-3gHt. 
\label{tile}\end{align}
This should be compared with the scaling and duality argument (\ref{QCdual}) 
and the one-loop quantum IR effect of Einstein gravity (\ref{Scaling1loop}), 
\begin{align}
V(f)=\exp(-\sqrt{\gamma}\kappa f)\sim 1-2\gamma Ht. 
\end{align} 
The above agreement by the both linear and exponential potentials implies that 
what we have proven with the exponential potential holds in the linear potential 
as well at the one-loop level.

In Sec. 4, this system is solved exactly by the Fokker-Planck equation for small $\gamma$,  
\begin{align}
g(t)=\frac{2}{\log(1+6Ht)+\frac{1}{g}}. 
\label{Resum}\end{align}
This one-loop exact solution can reproduce (\ref{tile}) at $Ht\ll 1$ and 
describe the global behavior at $Ht>1$. 
However, the scale of the inflaton is restricted to 
$\sqrt{\gamma}\kappa f \ll 1$. 
In other words, the deformation from de Sitter space has been evaluated as a linear response. 
In order to discuss the larger scale $\sqrt{\gamma}\kappa f>1$, 
we need to solve the Fokker-Planck equation for a time dependent background. 
Such an investigation is irrelevant with the current Universe (tiny $g$) 
while it is relevant with the primordial Universe (nonsmall $g$). 

For a time dependent $g(t)$, the left-hand side of the Fokker-Planck equation becomes as follows 
\begin{align}
\frac{1}{2}\frac{\partial}{\partial t}\big(\log \frac{\xi(t)}{g(t)}\big) \rho
-\frac{\partial}{\partial t}\big(\frac{4\xi(t)}{g(t)}\big)\omega^2\rho. 
\label{br1}\end{align}
In place of (\ref{Fpds}), we obtain 
\begin{align}
\frac{\partial}{\partial N}\log \frac{\xi}{g}=-6\xi, 
\label{xieqmrs}\end{align}
where the e-folding number $N$ is not exactly equal to $Ht$. 
Since $1/\xi$ measures the magnitude of the enhancement of the scalar perturbation (\ref{Stsol2}),
$\xi$ corresponds to the tensor-to-scalar ratio $r$. 

The exact solution of (\ref{br1}) is given by 
\begin{align}
g=\frac{2}{\log N}\big(1-\frac{1}{\log N}\big),\hspace{1em}
\xi=\frac{1}{6N}\big(1-\frac{1}{\log N}\big). 
\label{br2}\end{align}
The corresponding $\beta$ function can be evaluated as follows 
\begin{align}
\beta(g)=\frac{\partial}{\partial \log N}g=-\frac{2}{(\log N)^2}\big(1-\frac{2}{\log N}\big). 
\label{br3}\end{align}
In the IR region $N\gg 1$, (\ref{br2}) behaves similarly to (\ref{Resum})  
because the $1/\log N$ correction is tiny. 
Since $\epsilon$ decreases as $N$ increases, 
\begin{align}
\epsilon=-\frac{1}{2}\frac{\partial}{\partial N}\log g=\frac{1}{2N\log N}\big(1-\frac{1}{\log N-1}\big), 
\end{align}
the spacetime expansion does not become decelerating. 
It is a future subject to find a mechanism to end the inflation era. 

Given the $1/\log N$ correction, 
there exist not only the IR fixed point $g=0$ but also a UV fixed point. 
The $\beta$ function (\ref{br3}) shows that $g$ increases monotonically toward the past 
and has the maximum value $g=1/2$. 
The existence of the UV fixed point may indicate the consistency of quantum gravity 
and a conformal invariance in the beginning of the Universe. 
This situation is analogous to AdS/CFT. 
Near the IR fixed point, i.e., for the weak coupling, quantum Einstein gravity is a good approximation. 
On the other hand, the UV fixed point may indicate 
the existence of a strong coupled conformally invariant phase. 

We are caught by surprise to find that 
the exact $\beta$ function (\ref{br3}) possesses the UV fixed point in addition to the IR fixed point. 
Since we have adopted the Gaussian approximation, 
this is not a proof of the consistency of quantum gravity 
as the critical coupling $g=1/2$ is strong. 
Nevertheless, the messages are clear that quantum gravity on de Sitter space is both IR and UV finite.
The cosmic expansion started at the Planck scale with the minimal entropy $S=2$. 
We believe that this result underscores the holographic nature of gravity. 
We have investigated the degrees of freedom at the horizon. 
That is presumably all that matters as far as gravity is concerned. 

Let us consider the current accelerating Universe. 
The energy contents of the current Universe are given by 
the dark energy density $\Omega_{\Lambda}=0.7$ and the matter density $\Omega_M=0.3$. 
The incoming matter energy flux gives rise to the same phenomena, 
\begin{align}
\frac{-\dot{H}(t_0)}{H_0^2}
=\frac{4\pi G_N}{H_0^2} \rho_M
=\frac{4\pi G_N}{H_0^2}\frac{3H_0^2}{8\pi G_N}\Omega_M
=\frac{3}{2}\Omega_M\sim \frac{1}{2}, 
\label{dualarg}\end{align}
where $O_0$ denotes the present value of $O$. 
This effect is self-consistent since the current Universe is accelerating $\gamma=1/2 < 1$. 
Since this equation (\ref{dualarg}) follows from the Friedmann equation,
\begin{align}
\frac{H^2(t)}{H_0^2}= \frac{\Omega_M}{a^3}+\Omega_\Lambda, 
\label{Fried}\end{align}
it is hard to dispute. 
This Friedmann equation is a standard one and thus does not include new effects. 
We can estimate the time dependence of the Hubble parameter for small $z=-H_0(t-t_0)$, 
\begin{align}
\frac{H^2(t)}{H_0^2}
&=\Omega_M+\Omega_\Lambda+\Omega_M (e^{-3H_0t}-1) \notag\\
&\sim 1-0.9H_0(t-t_0). 
\label{DMHc}\end{align}

We recall the following relation between the energy density parameter $\Omega$ 
and the energy density $\rho$: 
\begin{align}
\Omega_M=g\rho_M\Theta\sim 0.3,\hspace{1em}
\Omega_{\Lambda}=g\rho_{\Lambda}\Theta\sim 0.7, 
\label{energyden}\end {align}
where $\Theta$ is the volume  of $S^4$ of radius $1/H_0$.
If the dark energy stays constant, it will be the sole 
energy component after the matter disappears, 
\begin{align}
\Omega_\Lambda=g_F\rho_\Lambda \Theta_F= 1, 
\end{align}
where $\Theta_F$ and $g_F$ are given by the final Hubble parameter $1/H_F$. 

In terms of the $\tau$ variable, it is manifest that we can scale the solution by $\tau\rightarrow c\tau$. 
By using this freedom, we may change the coupling $g$ into the energy density $\Omega$ 
by $g\rightarrow g\rho\Theta = \Omega$ using the relation in (\ref{energyden}). 
The point of this scaling is to effectively magnify minuscule $g$ into $\mathcal{O}(1)$ quantity $\Omega$. 
We can use the following expression for $\mathcal{O}(1)$ quantity: 
\begin{align}
\Omega(t)=\frac{2c^{-1}}{\log (1+Ht)}, 
\label{Omegaest}\end{align}
where $c$ is a normalization coefficient.
For small $t$, the above expression becomes, 
\begin{align}
\Omega(t)=\frac{2c^{-1}}{Ht}\sim\frac{2c^{-1}}{-H\tau}. 
\end{align}
So we obtain a condition $Ht >1$ for logarithmic behavior of $\Omega(t)$. 
It coincides when the Universe began the recent accelerated expansion. 
Instead of considering $\mathcal{O}(1)$ quantity $\Omega$, 
we can consider the ratio $H^2(t)/H_0^2$ which is certainly $\mathcal{O}(1)$. 
In fact, they are the same quantify $H^2(t)/H_0^2=\Omega (t)$. 

The quantum effects start to kick in at $t=t_i$. 
Dark energy may be no longer constant. 
Instead, it may initiate the logarithmic decay. 
Mathematically, it just sets the initial condition of the renormalization group for $g$. 
The initial condition is prepared by the classical Friedmann equation. 

We combine dark energy and matter effect (\ref{DMHc}), 
\begin{align}
\frac{H^2(z)}{H_0^2}=0.3(1+z)^3 +0.7\log \big(e+ \log(1+z)\big), 
\label{cmfl}\end{align}
where the redshift variable $1+z=1/a$ is introduced to compare with the observations. 
We also make use of the time reversal operation (\ref{Ivml}) 
since the redshift variable $z$ looks backward in time. 
The solution of the $\beta$ function is obtained 
by the following reparametrization $1+Ht \rightarrow e+\log (1+z)$. 
The factor $e$ is inserted in such a way that $\log (1+Ht_0)=1$ at present $z_0=0$. 

Since $\log x $ with the identification $x=\log (1+z)$ cannot be normalized around $x=0$, 
the shift of time is inevitable. 
We have fixed the time translation freedom as $x\rightarrow x+e$ 
such that $\log ( x+ e)$ is normalized when $x=0$. 
This time shift does not modify the energy density of matter 
because the normalization condition $\Omega_M(x=0)=0.3$ removes the extra factor due to this time shift.
In our convention, 
there is no free parameter here although we have ignored the nonlinear correlation 
between $\Omega_M$ and $\Omega_\Lambda$. 
The threshold effects at $t=t_i$ are also neglected. 
We hope to improve the Eq. (\ref{cmfl}) in these aspects. 
We compute $\gamma$ to judge the Universe is accelerating if $\gamma<1$, 
\begin{align}
(1+z)\frac{\partial}{\partial z}H^2(z)=2\gamma H^2(z), 
\label{esgam}\end{align}
\begin{align}
\frac{\gamma H^2(z)}{H_0^2}=(1+z)\frac{\partial}{\partial z}\frac{H^2(z)}{2H_0^2}
=\frac{1}{2}\big(0.9(1+z)^3+ 0.7\frac{1}{e+\log (1+ z)}\big). 
\end{align} 

We propose the following formula in predicting the future energy density parameters: 
\begin{align}
\frac{H^2(a)}{H_0^2}
=0.3\frac{1}{a^3}+0.7\frac{1}{\log (e+ \log a)}, 
\label{cmfl2}\end{align}
\begin{align}
\frac{\gamma H^2(a)}{H_0^2}=-a\frac{\partial}{\partial a}\frac{H^2(a)}{2H_0^2}
=\frac{1}{2} \big(0.9\frac{1}{a^3}+ 0.7\frac{1}{e+\log a}\frac{1}{\log ^2(e+\log a )}\big). 
\end{align}
At present  $a=1$ and $z=0$, $H^2$ and $\gamma$ agree in both formulas. 
The future oriented formula (\ref{cmfl2}) is smoothly connected to (\ref{cmfl}) at $a=1$ and $z=0$. 
This is due to the time reversal symmetry pointed out  in (\ref{trsm}).

We find a formula in the linear approximation which is valid for small $z$: 
\begin{align}
H(z)= H_0(1+0.58z), 
\label{Hcrs}\end{align}
which is the sum of matter effect and dark energy effect as $0.58=0.45+0.13$. 
We determine $z_i$ by requiring $\gamma (z_i)=1$: when the Universe started the accelerated expansion. 
It turns out to be $z_i=0.6$ and $\Omega_M =1.2$, $\Omega_\Lambda=0.8$. 
At present, $z=0$ and $\Omega_M=0.3$, $\Omega_\Lambda=0.7$, $\gamma_0=0.2$. 
This $\gamma_0$ is considerably smaller than the classical theory's one $\gamma_0=1/2$ in (\ref{dualarg}). 

The qualitative understanding of these numbers is easy in the classical case 
since dark energy density stays constant. 
Given the present energy density parameters, $\Omega_M(z)=(1+z)^3\Omega_M$, 
$\gamma$ is given by  $(3/2)\Omega_M(t)/(\Omega_M+\Omega_\Lambda)$. 
In the classical case, $\gamma=1$ corresponds to $\Omega_M(z)=4/3$ and $z=0.6$. 

In the quantum case where dark energy decays logarithmically, 
it turns out that $\Omega_M=1.2$, $\Omega_\Lambda=0.8$ when $\gamma=1$. 
The fact that $z_i=0.6$ comes out to be in the right ballpark is a nontrivial check 
of our theory against the observations. 

\begin{figure}%[t]
\begin{center}
\includegraphics[height=23pc]{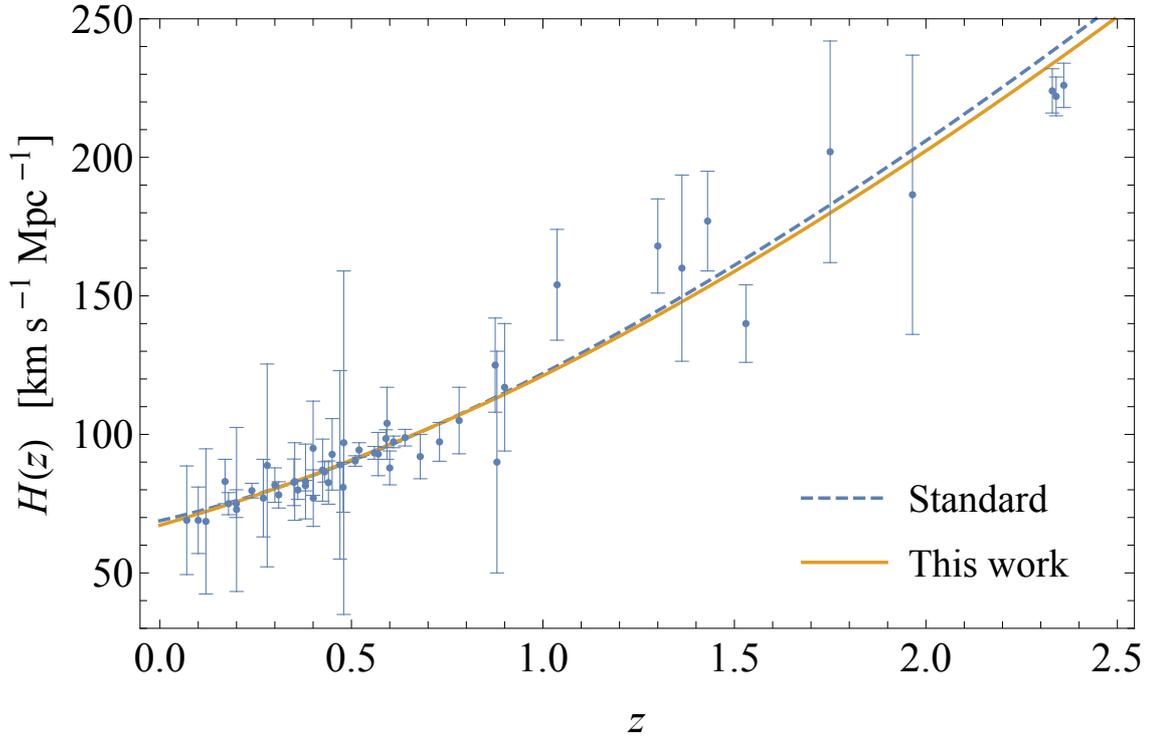}%{Hubble.eps}
\caption{\label{fig:Hubble}
The Hubble parameter measurements and their errors (in units of $\mathrm{km\ s^{-1}\ Mpc^{-1}}$) \cite{GZ}  
are compared with theoretical predictions.}
\end{center}
\end{figure}

In Fig.\ \ref{fig:Hubble}, the Hubble parameter measurements $H(z)$ 
and their errors $\sigma_H$ at 51 redshifts $z$ are plotted. 
The data are taken from the compilation of various observations in \cite{GZ}. 
For the theoretical curves, we fix the values of density parameters 
as $\Omega_M = 0.311$ and $\Omega_\Lambda = 0.689$ \cite{Planck}. 
However, we do not fix the Hubble constant $H_0$ 
because an observed value of $H_0 = 73.24 \pm 1.74\ \mathrm{km\ s^{-1}\ Mpc^{-1}}$ 
by type Ia Supernovae and Cepheids \cite{Rie} is systematically larger 
than an observed value of $H_0 = 67.66 \pm 0.42\ \mathrm{km\ s^{-1}\ Mpc^{-1}}$ by CMB \cite{Planck}. 
The origin of the discrepancy is not clear at the present time. 
For our model, we use
\begin{align}
\frac{H^2(z)}{H_0^2} =\Omega_M (1+z)^3 +
\begin{cases}
\Omega_\Lambda \log\big(e + \log (1+z)\big), & (z \leq 0.6) \\
\Omega_\Lambda \log\big(e + \log (1.6)\big), & (z > 0.6)
\end{cases}.
\label{eq:tm1}\end{align}

\begin{table}
\begin{center}
\begin{tabular}{cccc}
& $H_0\ [\mathrm{km\ s^{-1}\ Mpc^{-1}}]$
& $\chi^2/\mathrm{dof}$
%  & $p$-value
\\[.5pc] \hline\hline
Standard &
$68.8 \pm 2.5$
& $0.739$
%  & $0.39$
\\[.5pc] \hline
This work &
$67.2 \pm 2.5$
& $0.623$
%  & $0.43$
\\[.5pc] \hline
\end{tabular}
\caption{\label{tab:tm1}Chi-squares per degrees of freedom}
\end{center}
\end{table}
The results of the fitting are summarized in Table\ \ref{tab:tm1}. 
The figure-of-merit for the observed Hubble data is given by the chi-square per degrees of freedom, 
which is defined by
\begin{align}
\chi^2/\mathrm{dof}
=\frac{1}{N_\mathrm{dat}-N_\mathrm{par}}
\sum_{i=1}^{N_\mathrm{dat}}
\frac{\left[H_\mathrm{theo}(z_i) - H_\mathrm{obs}(z_i)\right]^2}{\sigma_{H,i}^2},
\label{eq:tm2}\end{align}
where $N_\mathrm{par}$ is the number of free parameters, 
$N_\mathrm{dat}$ is the number of the observational Hubble parameter $H_\mathrm{obs}(z_i)$ at redshift $z_i$, $\sigma_{H,i}$ is its error, and $H_\mathrm{theo}(z_i)$ is the theoretical value for a given model. 
In the present case, we have $N_\mathrm{par}=1$, $N_\mathrm{obs} = 51$. 
If the value of $\chi^2/\mathrm{dof}$ is much larger than unity, 
the assumed theory poorly explains the data. 

The differences between the standard model and our theory are not significant, 
and both can almost equally explain the observations of the Hubble parameters. 
The fitted value of the Hubble constant with our theory is slightly closer to the value estimated by CMB. 
The value of $\chi^2/\mathrm{dof}$ in our theory is slightly smaller than that in standard model. 
The difference is not significant, but this suggests that our theory has a slightly better fit 
than the standard model. 
One of the main reason comes from the fact that 
the data of highest redshifts around $z\sim 2.35$ are smaller than the expectations of the standard model, 
and our theory predicts smaller Hubble parameter for the high redshifts, 
because the decaying nature of the dark energy makes the slope of the curve shallower 
than the cosmological constant. 

Next, we consider the parameters of dark energy for $z < 0.6$. 
It is quite common to parametrize the equation of state of the dark energy 
by $w(a) = w_0 + w_a(1-a)$ \cite{CP,Lin}. 
Comparing the Eq. (\ref{cmfl}) with the standard theory, 
an integral $3\int_0^z dz(1+w)/(1+z)$ corresponds to the function $\log(\log(e+\log(1+z)))$ in our theory. 
The analytic solution is 
\begin{align}
3(1+w)= \frac{1}{\log(e+\log(1+z))}\frac{1}{e+\log(1+z)}. 
\end{align}
Applying Taylor expansions to both quantities and comparing the coefficients of $z^0$ and $z^1$, 
we obtain 
\begin{align}
w_0 = -1 + \frac{1}{3e},\hspace{1em}
w_a = -\frac{2}{3e^2}.
\label{eq:tm4}\end{align}
These are interesting predictions of our theory.

The observed values of $w_0$ and $w_a$ \cite{Ala}, 
predictions of standard cosmology and of our theory are respectively given by 
\begin{align}
&\mathrm{Observation:}
&&w_0 = -0.91 \pm 0.10,
&&w_a = -0.39 \pm 0.34, \label{eq:tm5a}\\
&\mathrm{Standard:}
&&w_0 = -1,
&&w_a = 0, \label{eq:tm5b}\\
&\mathrm{This\ work:}
&&w_0 = -0.877\cdots,
&&w_a = -0.090\cdots. \label{eq:tm5c}
\end{align}
While both predictions are consistent to the observation, 
the values of our theory are closer to the observed values 
than those of the standard cosmology. 
Thus our theory is promising, 
although more accurate observations will be necessary to judge 
if our theory is significantly better than the standard one or not. 

To conclude, there are some differences 
between our theory and the standard theory for the predictions of $H(z)$. 
The observed values of $H(z)$ and dark energy parameters $w_0$, $w_a$ are slightly closer 
to the predictions of our theory. 
However, the difference is not statistically significant at the error levels of the present time. 
Currently ongoing, and near-future observations such as DESI \cite{DESI} and Euclid \cite{Euclid} 
will significantly reduce the errors by factor $1/3$--$1/10$ for both parameters. 
Therefore, our theory will definitely be testable in near future when those observations are available. 

\section{Conclusions}
\setcounter{equation}{0}

We have investigated IR dynamics in quantum and classical gravitational theories 
on de Sitter-type space. 
We have formulated dynamical scaling law in generic four-dimensional gravitational theories. 
We have realized a duality between quantum IR effects in Einstein gravity 
and classical effects in an inflation (or quintessence) theory in four dimensions
just like in two dimensions \cite{KK2d}. 
Namely, quantum IR effects in Einstein gravity can be interpreted 
as classical phenomena in the inflation theory. 
As an example, the shielding of the Hubble parameter $H$ is found to occur in Einstein gravity 
due to the quantum diffusion of the conformal mode. 
We can identify the dual inflation theory in which $H$ decreases due to classical physics. 
The inflaton slowly rolls down the linear potential. 
The nontrivial point in this duality is that the inflaton potential is uniquely fixed. 
In fact, we find it necessary to introduce an inflaton into Einstein gravity 
as a counterterm to cancel the noncovariant IR logarithm. 
 
We thus postulate a duality between quantum Einstein gravity/a classical inflation theory. 
In our view, they may be the same thing seen from different angles. 
In the quantum gravity point of view, the main character is the conformal mode. 
The Hubble parameter decreases due to the stochastic process at the horizon. 
The de Sitter entropy is nothing but the von Neumann entropy of the conformal zero mode.
They increase due to diffusion at the horizon in a consistent way with the Gibbons-Hawking formula. 
In the dual picture, the Hubble parameter decreases due to the classical drift of the inflaton. 
The de Sitter entropy increases due to the incoming energy flow of the inflaton. 
So we can explain the same physics in terms of quantum effects in Einstein gravity 
and by classical physics in its dual inflation theory. 

This line of thinking puts strong constraints on possible outcomes of physics in the accelerating expanding eras. 
For example, we may be able to make unique predictions on the inflaton potential or the fate of dark energy. 
In this paper, we have evaluated the $\beta$ function of $g=G_N H^2/\pi$: 
the only dimensionless parameter in Einstein gravity. 
It turns out to be asymptotically free toward the future: $\beta(g)=(\partial/\partial \log( Ht)) g =-(1/2)g^2$. 
It predicts that dark energy decays logarithmically. 
The exciting prospect is that this prediction may be well observable. 
 
It is remarkable that the $\beta$ function does not contain a small parameter while $g$ is minuscule. 
Our Universe is situated very close to the fixed point $g\sim 0$ with a huge entropy $1/g$. 
The $\beta$ function explains why $g$ is destined to vanish logarithmically with time. 
We have gathered more evidence for our identification of the de Sitter entropy 
with the von Neumann entropy of the conformal zero mode. 
In fact, the $\beta$ function tells how fast the de Sitter entropy increases. 
It coincides with the increasing speed of the von Neumann entropy due to quantum diffusion at the horizon. 
If our prediction for dark energy is verified by observations, 
we are likely to have solved a major part of the cosmological constant problem: 
its destined fate and the mechanism of asymptotic freedom of $g$. 
However, there are still mysteries on its evolution process. 
Why the Universe started accelerated expansion now after the inflation ended, 
and just after the Universe was created? 
We certainly need more detailed understandings on the evolution of the Universe. 

In the case of inflation, the prediction of the slow-roll parameter in Einstein gravity 
and its dual is too small for CMB. 
In this work, the emergence of the linear potential at the one-loop level is demonstrated, at least locally. 
This result underscores our previous scenario 
where the slow-roll parameter $\epsilon$ grows into an observable value 
by quantum and classical effects \cite{KKEV}. 
It is possible that a desirable scenario which starts and ends the slow-roll inflation 
can be obtained by combining this work and the previous work. 

Concerning higher order corrections to the $\beta$ function, 
the Gibbons-Hawking formula is suggestive. 
If we assume that the inverse relationship $S=1/g$ holds to all orders qualitatively, 
the $\beta$ function must be negative on the whole way to the strong coupling limit 
since the entropy never decreases with time. 
In fact, such a point of view can be confirmed 
by the exact $\beta$ function within the Gaussian approximation (\ref{br3}). 
It is negative in the whole range of time flow. 
The surprising feature is that it has the UV fixed point in addition to the IR fixed point. 
The coupling approaches a finite value toward the past. 

Given the UV fixed point, it may be natural to assume 
the existence of a strongly coupled conformally invariant phase. 
Such an idea is old \cite{KKN} but the relevance of de Sitter space is a new insight. 
Physics may depend on the dimensionless coupling $g=G_NH^2/\pi$ only. 
In this combination, large $G_N$ is equivalent to large $H^2$. 
We need a nonperturbative framework to investigate such a possibility. 
Surprisingly, the IIB matrix model indicates that 
four-dimensional spacetime emerges out of matrices in de Sitter phase \cite{NT,BHM}.
It is serendipity that our work will be tested 
not only by observations but also by matrix models and string theory. 

\section*{Acknowledgment}
This work is supported by the National Center of Theoretical Sciences (NCTS), 
and Grant-in-Aid for Scientific Research (C) No. 16K05336, 
(B) No. 16H03977, and (C) No. 19K03835. 
We thank Chong-Sun Chu, Satoshi Iso, Hikaru Kawai, Kozunori Kohri, Yoji Koyama, Jun Nishimura,  
and Hirotaka Sugawara for discussions. 
We appreciate our referee's comment on higher order corrections to the $\beta$ function, 
which gave us the occasion to find the nontrivial UV fixed point. 

\appendix

\section{Gravitational propagators in de Sitter space}
\setcounter{equation}{0}

For self-containedness, we review gravitational  propagators in de Sitter space. 
The de Sitter background is given by 
\begin{align}
ds^2=a_c^2(-d\tau^2+dx_i^2),\hspace{1em}a_c=\frac{1}{-H\tau}.  
\end{align} 

The quadratic terms in the Einstein-Hilbert action are given by  
\begin{align}
&\frac{1}{\kappa^2}\int d^4x \sqrt{-g}[R-6H^2]\big|_2 \notag\\
=&\frac{1}{\kappa^2}\int d^4x \big[
-\frac{1}{4}a_c^2\partial_\mu h^{\rho\sigma}\partial^\mu h_{\rho\sigma} 
+\frac{1}{2}a_c^2\partial_\rho h^\rho_{\ \mu}\partial_\sigma h^{\sigma\mu} 
+2Ha_c^3h^{0\mu}\partial_\nu h^\nu_{\ \mu}
+3H^2a_c^4h^{0\mu}h^0_{\ \mu} \notag\\
&\hspace{4.5em}-2a_c^2\partial_\mu h^{\mu\nu}\partial_\nu\omega
-8Ha_c^3h^{0\mu}\partial_\mu\omega 
+6a_c^2\partial_\mu\omega\partial^\mu\omega
-24H^2a_c^4\omega^2 \big]. 
\label{EH}\end{align}
We adopt the following gauge fixing term: 
\begin{align}
\int d^4x\mathcal{L}_\text{GF}
&=\frac{1}{\kappa^2}\int d^4x\big[-\frac{1}{2}a_c^2F_\mu F^\mu\big], \notag\\
F_\mu&=\partial_\rho h^\rho_{\ \mu}-2\partial_\mu \omega
+2Ha_c h^0_{\ \mu}+4Ha_c\delta_\mu^{\ 0}\omega. 
\label{GF}\end{align}

The sum of (\ref{EH}) and (\ref{GF}) is given by 
\begin{align}
&\frac{1}{\kappa^2}\int d^4x\sqrt{-g}[R-6H^2]\big|_2+\int d^4x\mathcal{L}_\text{GF} \notag\\
=&\frac{1}{\kappa^2}\int d^4x\big[
a_c^2(-\frac{1}{4}\partial_\mu \tilde{h}^{ij}\partial^\mu \tilde{h}^{ij} 
+\frac{1}{2}\partial_\mu h^{0i}\partial^\mu h^{0i}
-\frac{1}{3}\partial_\mu h^{00}\partial^\mu h^{00} 
+4\partial_\mu\omega\partial^\mu\omega) \notag\\
&\hspace{4.5em}+H^2a_c^4(h^{0i}h^{0i}-h^{00}h^{00}+4h^{00}\omega-4\omega^2) \big], 
\end{align}
where $\tilde{h}^{ij}$ is the spatial traceless mode: 
\begin{align}
\tilde{h}^{ij}\equiv h^{ij}-\frac{1}{3}h^{00}\delta^{ij}. 
\end{align}
The quadratic action is diagonalized as follows 
\begin{align}
\frac{1}{\kappa^2}\int d^4x\big[
&-\frac{1}{4}a_c^2\partial_\mu \tilde{h}^{ij}\partial^\mu \tilde{h}^{ij} 
+\frac{1}{2}a_c^2\partial_\mu X\partial^\mu X \notag\\
&+\frac{1}{2}a_c^2\partial_\mu h^{0i}\partial^\mu h^{0i}+H^2a_c^4h^{0i}h^{0i} 
-\frac{1}{2}a_c^2\partial_\mu Y\partial^\mu Y-H^2a_c^4 Y^2 \big], 
\label{diagonalized}\end{align}
where $X$ and $Y$ are given by 
\begin{align}
X\equiv 2\sqrt{3}\omega-\frac{1}{\sqrt{3}}h^{00},\hspace{1em}Y\equiv h^{00}-2\omega. 
\end{align}
Furthermore, the quadratic Fadeev-Popov ghost term is given by 
\begin{align}
\int d^4x\mathcal{L}_\text{FP}\big|_2= \frac{1}{\kappa^2}\int d^4x
\big[-a_c^2\partial_\mu \bar{b}^i\partial^\mu b^i
+a_c^2\partial_\mu \bar{b}^0\partial^\mu b^0+2H^2a_c^4\bar{b}^0b^0\big]. 
\label{FP}\end{align}

As seen in (\ref{diagonalized}) and (\ref{FP}), 
Einstein gravity consists of massless minimally coupled modes, 
and conformally coupled modes. 
We neglect the conformally coupled modes 
\begin{align}
h^{0i}\simeq 0,\hspace{1em}Y\simeq 0,\hspace{1em}b^0\simeq 0, 
\end{align}
and focus on the subspace of massless minimally coupled modes 
\begin{align}
h^{00}\simeq 2\omega \simeq \frac{\sqrt{3}}{2}X,\hspace{1em}\tilde{h}^{ij},\hspace{1em}b^i. 
\end{align}
That is because only the massless minimally coupled modes induce the IR logarithms
\begin{align}
\langle X(x)X(x')\rangle&=-\langle \varphi(x)\varphi(x')\rangle, \notag\\
\langle \tilde{h}^{ij}(x)\tilde{h}^{kl}\rangle
&=(\delta^{ik}\delta^{jl}+\delta^{il}\delta^{jk}-\frac{2}{3}\delta^{ij}\delta^{kl})\langle \varphi(x)\varphi(x')\rangle, 
\notag\\
\langle b^i(x)\bar{b}^j(x')\rangle&=\delta^{ij}\langle \varphi(x)\varphi(x')\rangle, 
\end{align}
where the two-point function of a massless minimally coupled scalar field is given by 
\begin{align}
\langle \varphi(x)\varphi(x')\rangle\simeq \frac{\kappa^2H^2}{8\pi^2}\log \big(a_c(\tau)a_c(\tau')\big). 
\end{align}
As discussed in the main text, the negative norm of $X$ (i.e., $h^{00}$ and $\omega$) plays an important role 
to screen the cosmological constant. 

\section{One-loop IR logarithms and duality}
\setcounter{equation}{0}

We explain our investigations on the IR renormalization problem of Einstein gravity 
from the duality point of view. 
Although we can work in any conformal frame, 
we pick the following frame where the background $a$ is the classical solution. 
Sometimes, we find it convenient to assume $a$ is slightly off-shell, 
\begin{align}
\frac{1}{\kappa^2}\int d^4x \big[&a^2\phi^2\tilde{R}
-6a\phi\partial_\mu\big\{\tilde{g}^{\mu\nu}\partial_\nu(a\phi)\big\}
-6H^2a^4\phi^4\big], 
\label{CFR}\end{align}
where we parametrize $\phi=e^{\omega}$. 
The quantum equation is the condition that there is no tadpole. 
In our case, it is equivalent to require that the coefficient in front of $\omega,\ h^{00}$ must vanish. 
In other words, there should be no linear term in $\omega,\ h^{00}$ in the effective action. 
Since the de Sitter solution satisfies this requirement, 
the classical action corresponds to $\omega=h^{00}=0$, 
\begin{align}
\frac{1}{\kappa^2}\int d^4x \big[\sqrt{-\hat{g}}(\hat{R}-6H^2)
=6a\partial_0^2a-6H^2 a^4\big]. 
\label{ClAct1}\end{align}
 
There may be a gauge and a parametrization where IR logarithmic effects are suppressed 
by derivative interactions. 
We perform partial integrations a few times to find such a condition. 
We suppress the $\partial Z\partial Z$-type term ($Z$ denotes $h^{\mu\nu}$ or $\omega$) in what follows. 
Such candidates are listed below 
\begin{align}
\frac{1}{\kappa^2}\int  d^4x \big[
2\partial_0a^2\tilde{g}^{0\nu}\partial_\nu\phi^2
+(6\partial_0a\partial_0a-\partial_0^2a^2)\tilde{g}^{00}\phi^2
-6H^2a^4\phi^4 \big], 
\label{DIR4}\end{align}
\begin{align}
\frac{1}{\kappa^2}\int d^4x \big[
-2\partial_0a^2 \partial_\nu\tilde{g}^{0\nu}\phi^2-a^4\hat{R}\tilde{g}^{00}\phi^2
-6H^2a^4\phi^4 \big]. 
\label{DIR5}\end{align}
The former (\ref{DIR4}) shows the equation of motion with respect to $h^{00}$ 
and the equation of motion with respect to $\phi$ is manifest in the latter (\ref{DIR5}) respectively. 

After these preparations, we integrate the IR fluctuations around the classical solution. 
The quantum equation at the one-loop level requires that no field comes out of the loop. 
So we need three-point vertices. 
The gauge fixing term is necessary only to define propagators. 
We use the exponential parametrization of the metric and the quadratic gauge fixing term. 
After diagonalization, we have a massless minimally coupled mode $X$ and a conformally coupled mode $Y$. 
The latter does not have the large IR fluctuation unlike the former. 
We regard it to be constant sitting at the minimum of the potential. 
The other modes do not contribute the IR logarithms to the cosmological constant. 
We decompose $h^{00}$ and $2\omega$ into
\begin{align}
h^{00}=AX+3BY,\hspace{1em}2\omega= AX+BY,\hspace{1em}
(A,B)=(\frac{\sqrt{3}}{2},\frac{1}{2}). 
\label{Sfdc}\end{align}
We need to calculate the one-point function of $\omega,\ h^{00}$ 
or take a derivative of the effective action with respect to $\omega,\ h^{00}$.
Since we are interested in IR logarithms, we can identify $h^{00}=2\omega$ for internal loop. 

We focus on a singly differentiated term in (\ref{DIR5}): 
\begin{align}
&-2\partial_0a^2\partial_\nu\tilde{g}^{0\nu}\phi^2
=2\partial_0a^2\partial_0 e^{h^{00}} e^{2\omega} \notag\\
=&2\partial_0 a^2\partial_0 e^{(AX+3BY)} e^{(AX+BY)}
=\partial_0a^2\partial_0e^{2AX+4BY}, 
\end{align}
where we assume $Y$ is constant. 
Therefore, we obtain 
\begin{align}
\int d^4x \big[\partial_0a^2\partial_0 e^{2AX}e^{4BY}
=-\partial_0^2a^2 e^{2AX}e^{4BY}
=-\partial_0^2a^2 e^{h^{00}}\phi^2\big]. 
\end{align}

We adopt the approximation $h^{00}=2\omega$ which is valid in the subspace 
with massless minimally coupled modes and the gauge we have chosen. 
It is because $h^{00}$ and $2\omega$ can be identified with the massless minimally coupled scalar field $X$.  Nevertheless, the original composition of the operators should enable us
to identify them as $\sqrt{-g}R$ or $\sqrt{-g}$. 

In this way, we obtain the interaction potential: 
\begin{align}
\frac{1}{\kappa^2}\int  d^4x \big[
&\frac{1}{2}\big\{a^4\hat{R}-(6\partial_0a\partial_0a-\partial_0^2a^2)\big\}e^{h^{00}}\phi^2
-6H^2a^4\phi^4 \notag\\
&=\big\{a^4\hat{R}-(6\partial_0a\partial_0a-\partial_0^2a^2)\big\}e^{2AX}e^{4BY}
-6H^2a^4e^{2AX}e^{2BY} \big]. 
\label{DIR6}\end{align}
We can perform the same approximation in (\ref{DIR4}) with the identical result as the above. 
Note that potential vanishes on-shell in $X$ field space. 
By evaluating the expectation value of the two-point functions of the interaction potential, 
we reproduce (\ref{1lpeff3}) in Sec. 3.

We briefly recall our renormalization prescription given in Sec. 3.
In order to prepare the counterterm, 
we perform the conformal transformation $a\rightarrow a a_c^\gamma$. 
We introduce an inflaton $f$ and its potential $\exp(-2\Gamma f)$ 
to subtract the noncovariant IR logarithm by a covariant counterterm. 
The inflaton potential is chosen to let the classical solutions of the conformal mode and the inflaton coincide. 
Since it is equal to $a_c^{-2\gamma}$, it undoes the half of the conformal transformation 
of $H^2a^4\rightarrow H^2a^4a_c^{4\gamma}$. 
The remaining  overall $a_c^{2\gamma}$ factor acts as the counterterm for $\kappa^2$. 
This constitutes our counterterm (\ref{DIR6m}) to the one-loop quantum correction (\ref{1loopefac}). 
By combining them, we reproduce the one-loop effective action (\ref{DIR7m}) 
and the solutions in the background gauge (\ref{Betafnt}): 
\begin{align}
H^2(t)&=H^2\big(1-\frac{3}{4}\frac{\kappa^2H^2}{4\pi^2}\log a_c\big), \notag\\
a^2&=a_c^2\big(1+\frac{3}{4}\frac{\kappa^2H^2}{4\pi^2}\log a_c\big), \notag\\
\kappa^2&= \text{const}. 
\label{Qeqsol}\end{align}
These results imply the essence of our approximation is to restrict the field space to that of $X$ field. 

This scaling solution is not an exact solution of the quantum equation. 
Since the scalar curvature is not covariant under the conformal transformation, 
we still need to deal with the following drift force, 
\begin{align}
V'=12\gamma\big\{\partial_0^2 (a\log a_c)-(\partial_0^2 a)\log a_c\big\}. 
\end{align}
We may deform the solution (\ref{Qeqsol}) to balance the tree and quantum effects, 
\begin{align}
12\partial_0^2 \delta a-72H^2a^2\delta a+V'=0\ \rightarrow\ -24\partial_0^2 \delta a+V'=0. 
\end{align}
The new solution is
\begin{align}
a= a_c\big\{1+\frac{3}{8}\frac{\kappa^2H^2}{4\pi^2}(\log a_c+\frac{3}{4})\big\}. 
\end{align}

We may alternatively perform finite renormalization of the Einstein-Hilbert action 
to eliminate the finite correction in $a$ as follows 
\begin{align}
\frac{1}{\kappa^2}\int d^4x \big[&\sqrt{-g}\big\{R-2(1+\gamma)(3+2\gamma)H^2(t)\big\} \notag\\
&\to (6-2\gamma)a\partial_0^2a-2(1+\gamma)(3+2\gamma)H^2(t) a^4 \big], 
\label{ClAct2}\end{align}
to keep $a= a_c(1+\frac{3}{8}\frac{\kappa^2H^2}{4\pi^2}\log a_c)\sim a_c^{1+\gamma}$.\footnote{
Recall that $a_c=1/(-H_R\tau)$, $H_R^2=H^2(1+\frac{3}{2}\gamma)$. }

This solution must solve the equation of motion with respect to $h^{00}$ (\ref{SGIF}) 
since it exhibits the dynamical scaling law. 
The Hubble parameter is found to scale with the size of the Universe 
while we fix the Newton's coupling to be constant. 
In order to match finite terms, we need to renormalize the scalar curvature operator as above: 
\begin{align}
\sqrt{-g}R = a^2\tilde{R} +6\tilde{g}^{\mu\nu}\partial_\mu a\partial_\nu a
\rightarrow a^2\tilde{R} +(6-2\gamma)\tilde{g}^{\mu\nu}\partial_\mu a\partial_\nu a. 
\end{align}
The purpose of this appendix is to convey the power of duality. 
The quantum solution with a nontrivial scaling exponent $\gamma$ in Einstein gravity 
can be constructed as a classical solution of the dual inflation theory (\ref{Inflaton}). 
The fulfillment of the equation of motion with respect to $h^{00}$ and general covariance are manifest 
in the dual inflation theory 
while they are secretly hidden in the effective action of Einstein gravity (\ref{CMact}).

\end{document}